\documentstyle[12pt,aaspp4,epsf]{article}
\begin{document}

\def\intldate{\number\day\space\ifcase\month\or
January\or February\or March\or April\or May\or June\or
July\or August\or September\or October\or November\or December\fi
\space\number\year}


\def \ncite  #1{#1\ --{\it get ref}}
\def \deg    {$^{\circ}$}
\def \etal   {{et al.\thinspace}}
\def \eg     {{e.g.,}}
\def \cf     {{cf.}}
\def \ie     {{i.e.,}}
\def \hub    {$H_{\hbox{\rm 0}}$}
\def \hunits {km s$^{\hbox{\rm --1}}$ Mpc$^{\hbox{\rm --1}}$}
\def \kms    {{\rm km~s$^{\hbox{\rm --1}}$}}
\def \sec    {$^{s}$}
\def \arcsecpoint {$^{''}\mkern-5mu.$}
\def \asec	{\arcsecpoint}
\def \dophot	{D{\sc o}PHOT}
\def \DOPHOT	{D{\sc o}PHOT}

\def \hi    {\ion{H}{1}}
\def \tightenlines {\def\baselinestretch{1}\small}
\def\gtorder	{\mathrel{\raise.3ex\hbox{$>$}\mkern-14mu\lower0.6ex\hbox{$\sim$}}}
\def\ltorder	{\mathrel{\raise.3ex\hbox{$<$}\mkern-14mu\lower0.6ex\hbox{$\sim$}}}
\def\sb		{{\rm mag~arcsec$^{-2}$}}
\def\area	{${\rm deg}^2$}
\def\kpc	{\hbox{\rm kpc }}
\def\pc		{\hbox{ pc }}
\def\yr		{ \, {\rm yr}}
\def\peryr	{ \, {\rm yr^{-1} }}
\def\vlos	{ v_{\rm los} }
\def\lsim	{ \rlap{\lower .5ex \hbox{$\sim$} }{\raise .4ex \hbox{$<$} } }
\def\gsim	{ \rlap{\lower .5ex \hbox{$\sim$} }{\raise .4ex \hbox{$>$} } }
\def\solar	{ {\odot} }
\def\lsolar	{ {\rm L_{\odot}} }
\def\msolar	{ \rm {M_{\odot}} }
\def\surfmunit  { \rm {\, \msolar \, pc^{-2}} }
\def\HI		{{H{\sc I}}}
\def\etal	{{\it et~al.}}
\def\mags	{{ \, \rm mag }}   
\def\percubicpc	{ { \pc^{-3} } }
\def\abs	{ \hbox{ \vrule height .8em depth .4em width .6pt } \,} 
%
%
%

\title{The 1995 Pilot Campaign of PLANET:\\
\vskip 0.1cm
Searching for Microlensing Anomalies through 
Precise, Rapid, Round-the-Clock Monitoring}

\vskip 0.2cm

\author{M. Albrow\altaffilmark{1,2}, J.-P. Beaulieu\altaffilmark{3},
P. Birch\altaffilmark{4}, J. A. R. Caldwell\altaffilmark{1}, 
S. Kane\altaffilmark{5,6}, R. Martin\altaffilmark{4}, \\ 
J. Menzies\altaffilmark{1}, 
R. M. Naber\altaffilmark{3}, J.-W. Pel\altaffilmark{3},  
K. Pollard\altaffilmark{1}, P. D. Sackett\altaffilmark{3}, 
K. C. Sahu\altaffilmark{6}, \\ 
P. Vreeswijk\altaffilmark{3}, 
A. Williams\altaffilmark{4}, M. A. Zwaan\altaffilmark{3}}
\author{The PLANET Collaboration}
\affil{}

\altaffiltext{1}{South African Astronomical Observatory, P.O. Box 9, 
Observatory 7935, South Africa}
\altaffiltext{2}{Univ. of Canterbury, Dept. of Physics \& Astronomy, 
Private Bag 4800, Christchurch, New Zealand}
\altaffiltext{3}{Kapteyn Astronomical Institute, Postbus 800, 
9700 AV Groningen, The Netherlands}
\altaffiltext{4}{Perth Observatory, Walnut Road, Bickley, Perth~~6076, Australia}
\altaffiltext{5}{Univ. of Tasmania, Physics Dept., G.P.O. 252C, 
Hobart, Tasmania~~7001, Australia}
\altaffiltext{6}{Space Telescope Science Institute, 3700 San Martin Drive, 
Baltimore, MD. 21218~~U.S.A.}



\vskip 0.2cm

\centerline{\bf ABSTRACT}

{\tightenlines
PLANET (the Probing Lensing Anomalies NETwork) is a worldwide 
collaboration of astronomers whose primary goal is to monitor  
microlensing events densely and precisely 
in order to detect and study anomalies  
that contain information about Galactic lenses and sources 
that would otherwise be unobtainable.   
The results of PLANET's highly successful first year of operation 
are presented here. 
Details of the observational setup, observing procedures, and 
data reduction procedures used to track the progress in real time 
at the three participating 
observing sites in 1995 are discussed.  
The ability to follow several events simultaneously with a median 
sampling interval of 1.6 hours and a photometric precision of better 
than 0.10mag even at I=19 has been clearly demonstrated.  
During PLANET's 1995 pilot campaign, 10 microlensing events 
were monitored; the binary nature of one of these, MACHO~95-BLG-12 
was recognized by PLANET on the mountain.  Another event, 
OGLE~95-BLG-04, displayed chromaticity that may betray the presence 
of blending with unresolved stars projected onto the same resolution 
element.  Although lasting only about a month, the campaign 
may allow constraints to be placed on the number of planets 
with mass ratios to the parent star of 0.01 or greater.
}



\centerline{Accepted for publication in {\it The Astrophysical Journal\/} }

\newpage


\section{Introduction} \label{intro}

The phenomenon of gravitational microlensing, the apparent brightening 
and subsequent dimming of a background star as the gravitational field 
of a moving foreground star or other object alters the light path 
of the background source, was predicted by Einstein in 1936 (\cite{einstein36}).
Due to the precise alignment required for a detectable brightening, 
the chance of a substantial microlensing magnification is extremely small 
--- on the order of $1 \times 10^{-6}$ for background stars in the 
Galactic Bulge or nearby Magellanic Clouds, even if all the 
unseen Galactic dark matter is composed of objects capable of 
lensing (\cite{paczynski86}).  
For this reason, it was not until 1993, when massive observational programs 
capable of surveying millions of stars were well underway,  
that microlensing was observed toward the Galactic Bulge and Magellanic Clouds 
by the EROS, MACHO, and OGLE projects 
(\cite{aubourg93}, \cite{alcock93}, \cite{udalski93}).  
Gravitational microlensing toward the Galactic Bulge and the Large 
Magellanic Cloud (LMC) has now been convincingly demonstrated by the 
more than 200 microlensing events detected by these microlensing 
survey teams.  Other groups have joined the hunt by adding more events 
in the Bulge (\cite{alard95eso}) and 
candidate events in the nearby spiral M31 
(\cite{crottstom96m31}), while still others are beginning to collect 
data in the LMC (\cite{moa97}) and M31 (\cite{agape97}).  
Microlensing is providing new information on the amount of mass (dark and 
luminous) along these lines of sight (\cite{paczynski96araa}),
although the nature of the unseen lenses is still a matter of 
considerable debate and accurate interpretation will require more data 
(\cite{maopac96}). 
Reviews on the subject of microlensing can be found 
in \cite{paczynski96araa} and \cite{gould96pasp}.

The field of microlensing has quickly matured to the point that events are now  
reliably detected and reported while they are still in progress; the 
OGLE and MACHO teams have issued over 150 
real-time electronic ``alerts,'' nearly all in the direction of the 
Galactic Bulge (\cite{pratt96}, \cite{udalski94}).  
This capability has stimulated the birth of second-generation projects 
such as PLANET (\cite{albrow96}, \cite{albrow97}) 
and GMAN (\cite{pratt96})  
that gather detailed photometric information 
about individual events, producing a refined understanding of 
the primary light curve.  Other groups dedicated to obtaining 
high-quality spectroscopy of microlensing events 
are also in place (\cite{lennon97}). 
Most importantly, microlensing monitoring data of sufficient precision and  
quantity allows the detection and characterization of 
microlensing {\it anomalies\/}, fine structure 
in the light curves that contain detailed information about 
the nature of source and lens populations.  
As a result, new fields of study are 
just beginning to open as microlensing monitoring gives astronomy 
a tool for the study of the kinematics of lenses, 
the stellar mass function, 
the frequency and nature of binary systems, stellar atmospheres,   
and the detection and characterization of brown dwarfs 
and extra-solar planets. 

Galactic microlensing survey experiments are optimized to maximize  
the detection rate of microlensing events, but 
are ill-suited to the detection and characterization 
of most anomalies, necessitating a separate monitoring effort.  
The PLANET (Probing Lensing Anomalies NETwork) collaboration 
uses a worldwide network of telescopes to obtain the  
frequent, precise observations required for study of these anomalies.  
In particular, PLANET observations are optimized for the detection of 
anomalies caused by planets orbiting distant Galactic lenses, a subject that 
has come to the fore of scientific and public attention recently 
with the apparent detection via other techniques of nearby 
extra-solar planets.  
Described here are the results of its 1995 pilot campaign, during which 
10 microlensing events were observed for approximately four weeks 
from three semi-dedicated locations in the southern hemisphere. 
The campaign resulted in the most precise and densely-sampled microlensing 
light curves to date and the real-time detection of one binary system. 

A summary of microlensing anomalies and the information 
that they contain is given in \S\ref{anomalies}.  The PLANET collaboration is 
discussed in \S\ref{collab}, and the details of its 1995 pilot season are 
given in \S\ref{campaign}.  A description of the data reduction 
procedures appears in \S\ref{reduction}, and a discussion of the photometric errors follows in \S\ref{errors}. 
The resulting microlensing light curves are presented and discussed 
in \S\ref{results}.  Conclusions and final remarks can be found in 
\S\ref{conclude}.


\section{Microlensing Anomalies} \label{anomalies}

The ability of the survey teams to find microlensing needles in 
the dense, stellar haystacks of the Galactic Bulge and LMC has relied 
most heavily on the simple, symmetric, achromatic and non-repeating  
light curve that distinguishes simple microlensing from 
other variable phenomena.  
The form of this light curve, which is appropriate to lensing 
geometries in which both the point-source and point-lens approximations 
are valid and all motions are rectilinear, can be characterized by 
four parameters: the maximum magnification A, the time of maximum 
magnification $t_o$, the baseline flux $F_o$, and a characteristic 
width of the light curve $t_E$.  

The angular ``range of influence'' of a point-lens is characterized by 
its Einstein ring radius $\theta_E$, which is a 
function of the lens mass $M$ and the lens-source geometry:  
\begin{equation} \label{eringeq}
\theta_E = \sqrt{\frac{4 G M D_{LS}}{c^2 D_{L} D_{S}}},
\end{equation}
\noindent 
where $D_{S}$ is the observer-source distance, $D_{L}$ the observer-lens 
distance, and $D_{LS}$ the source-lens distance.  
Galactic microlensing by stars or other objects of similar mass 
generates two images separated by a distance equal to or greater  
than $2 \theta_E$, which for typical geometries and masses is on 
the order of 1 milliarcsecond and thus too small to be resolved 
by conventional techniques.   

The combined time-variable flux $F(t)$ of the two microimages, on 
the other hand, is detectable, and is given by 
\begin{equation} \label{feq}
F(t) = A(t) F_o 
\end{equation} 
\noindent 
where $F_o$ is the unlensed (baseline) source flux and A(t) is the 
changing microlensing magnification due to the relative motions 
of the observer, lens and source. 
For a point-source separated by an angular distance $\theta_{LS}$ at time $t$ 
from a point-lens, the magnification $A(t)$ is a simple function of 
the normalized angular separation $u(t) \equiv \theta_{LS}/\theta_E$: 
\begin{equation} \label{aeq}
A(t) = \frac{u^2(t) + 2}{u(t) \sqrt{u^2(t) + 4}}.
\end{equation}
\noindent 
Rectilinear motion of the observer-lens-source geometry in which 
the lens moves with speed $v_{\perp}$ across  
the observer-source line of sight  
results in a time dependence for $u(t)$ given by 
\begin{equation} \label{ueq}
u(t) = \sqrt{\frac{(t - t_o)^2}{t_E} + u^2_{min}},
\end{equation}
\noindent 
where $t_o$ is the time at peak magnification (when $u(t)$ attains its 
minimum value $u_{min}$) 
and $t_E \equiv  D_{L} \theta_E / v_{\perp}$ is the 
time required for the source to cross an Einstein ring radius. 

For a given population of lenses, reasonable estimates can be made 
for the amplitude and duration of microlensing light curves.   
Sources lying inside the Einstein ring in projection 
(\ie\ with $\theta_{LS} < \theta_E$ and thus $u_{min} < 1$) will have 
peak magnifications $A$ in excess of 1.34.  
Statistically, the peak magnifications  
will be distributed linearly in $u_{min}$, with smaller $u_{min}$ resulting 
in higher magnification $A$ and approaching $A = 1/u_{min}$ as 
$u_{min} \rightarrow 0$. 
A source in the Galactic Bulge moving with a speed of 200~\kms\ 
with respect to a $1 \msolar$ lens located half way between the 
source and the observer will have an Einstein time $t_E$ of $\sim$35 days. 
Due to the form of the stellar initial mass function and the 
distribution of Galactic light, 
typical stellar lenses are probably less massive and located closer to the 
Galactic center, resulting in a somewhat smaller typical $t_E$.  
The length of time that a typical Galactic event is above the canonical 
$A=1.34$ would be expected to be less than $2 t_E$ and thus on the order 
of weeks to months, roughly matching that observed (\cite{alcock97bulge1yr}). 

Microlensing anomalies are departures from the achromatic light curve 
given by Eqs.~\ref{feq}-\ref{ueq}.  Such departures are 
expected in the case of multiple lenses, multiple point sources or blends 
along the line-of-sight, extended sources, and complicated relative motion 
within the source-lens-observer system. 
Thus anomalies --- if well characterized --- can 
be used to extract detailed information about source and lens 
populations.  Such information is sorely needed in the interpretation 
of the primary survey data since of the four standard microlensing 
parameters, three contain 
no information about the lens, and the fourth, the characteristic 
Einstein time $t_E$, is a degenerate combination of the lens mass, 
distance, and relative velocity.   Examples of microlensing anomalies 
and the extra science that they can provide are detailed below.

\subsection{Lensing Binary Systems}

Multiple lenses separated by a distance up to a few Einstein ring radii 
no longer behave as isolated lenses.  
The axial symmetry of the magnification pattern is destroyed and 
non-linear effects generate caustic structure projected onto the 
source plane. 
Sources crossing caustic curves will exhibit sharp enhancements 
in their light curves with durations of a few hours, and while inside 
the curve the magnification will remain elevated due to the 
generation of additional images.  The magnification pattern outside 
the caustic curve is also distorted relative to that in the single 
point-lens case, and thus detectable light curve anomalies with a 
wealth of morphologies can be 
generated by multiple lens systems even if no caustic crossings occur. 
The OGLE group made the first clear detection of a binary lens in 1993 (\cite{oglebinary}) 
and since then several more have been reported 
(\cite{alard95aa}, \cite{alcock97bulge1yr}). 
Modeling of the well-sampled binary light curves yields the binary mass ratio 
and the binary separation in units of $\theta_E$.  Such events 
thus contain information about lensing binaries 
too faint to be detected by other methods in the disk and bulge of 
the Galaxy (\cite{domhir94}, \cite{maostef95}, \cite{gaudi97binaries}, 
\cite{dom98rotbin}).  

\subsection{Lensing Planetary Systems}

A special case of a multiple lens is a star with a planetary system.  
For certain geometries, a planet separated from its primary 
by an angular distance 
comparable to $\theta_E$ of the primary lens can  
create dramatic sharp peaks in the light curve with durations 
of a few hours to a few days (\cite{maopac91}), as long as 
the caustic structure does not resolve the source too severely.   
Photometry sensitive to 4-5\% deviations  
would result in detection efficiencies of $\gtorder$15-20\% 
for Jupiter-mass planets in the ``lensing zone,'' \ie\ 
Jupiters with projected angular  
separations (in units of the Einstein ring radius) of 
$0.6 \ltorder b \ltorder 1.6$ 
(\cite{gouldloeb92}, \cite{bolatto94}).  
Well-monitored events of very high magnification  
($A \, \gsim \,10$, corresponding to $u_{min} \, \lsim \, 0.1$) 
have detection sensitivities to anomalies caused by 
planets anywhere in the lensing zone that  
approach 100\% (\cite{griestsafi98}); 
such events are also more likely to betray the presence of 
multiple planets (\cite{gns98}).   
Since the lensing zones of stellar lenses on the line of sight to the 
Galactic Bulge are expected to lie between about 1 and 6 AU,  
and a large fraction of Galactic lenses are believed to be stellar 
(\cite{paczynski94}, \cite{zhao95}, \cite{hangould96}), 
microlensing is ideally suited to search for planetary systems like 
our own orbiting stars several kiloparsecs distant.
If events are monitored densely for a few Einstein times $t_E$, 
planets outside the lensing zone may also be detected via microlensing 
(\cite{stefscal97}).  

If extra-solar planetary systems are common, precise microlensing 
monitoring can produce distributions of planet-lens mass ratios and 
projected orbital radii, whereas non-detection of planetary lensing anomalies 
would place strong constraints on the numbers of Galactic stars with planets 
of mass greater than that of Neptune orbiting within several AU.  
Since the set of light curve morphologies for lensing 
planetary systems is large and varied (\cite{maopac91}, \cite{wambs97}),
planetary parameters will be degenerate unless densely-sampled, 
multi-band observations are available (\cite{gaudi97planets}).  
The caustic structure of smaller mass planets would 
resolve most stellar sources, resulting in a severe reduction in the size and 
chance of a planetary perturbation (\cite{bennettrhie96}).
Reliable detection of Earth-mass planets will thus require characterization of 
1-2\% deviations against non-giant sources for hundreds of events.  

Microlensing is thus a statistical technique for the detection and 
characterization of planetary systems that is sensitive to a 
large range in planetary mass, orbital characteristics and position 
in the Galaxy.  As such, it is complementary to radial velocity, 
pulsar timing, astrometry, and direct imaging methods which are 
designed for prolonged and detailed studies of individual objects.  
Although the field is still young, several reviews on the use of microlensing 
for the detection of extra-solar planets can be found in the literature 
(\cite{paczynski96araa}, \cite{peale97}, \cite{sackett97esorep}, 
\cite{sahu97}).

\subsection{Chromatic Anomalies: Binary Sources, Blends, and Resolved Sources}

Since a gravitational field deflects light of all wavelengths identically, 
one of the most fundamental characteristics of microlensing is the 
achromaticity of the resultant light curve.  Nevertheless, color variations, 
or chromaticity, can occur if the light is generated by anything other 
than a simple point source.  Binary (unresolved) source stars can  
create anomalies in light curve shape and chromaticity as first one 
and then the other of the pair is lensed 
(\cite{griesthu92bin}, \cite{hangould97}). 
A blend of a single source star with unassociated foreground 
star(s) can also create an anomaly (\cite{stefesin95}), 
which can be easily described 
by modifying Eq.~\ref{feq} to include 
an additional term $B$ to account for the sum of all flux from 
{\it unlensed\/} stars that perchance lie at small enough projected 
distances that they are unresolved.  Since these stars will in general 
have a different color than the lensed source, the microlensing 
light curve will have a color closer to that of the lensed star at 
peak, returning to the color of the average combined flux at baseline.
Blends are important to characterize since errors in estimates of 
the event time scales and survey sensitivities will otherwise occur 
(\cite{stefesin95}, \cite{alard97blends}, \cite{wozpac97}, \cite{han97}).  
Any light emitted by the lens itself will also cause 
blending (\cite{nemi97}), and since a large fraction of Galactic lenses may be 
stellar (\cite{paczynski94}, \cite{sahu94}, \cite{zhao95}, \cite{hangould96}), 
this may occur with some frequency.  
It has been suggested that measuring the frequency and strength of 
blending as a function of event duration could be used to obtain 
constraints on the mass of stellar lenses (\cite{buchalter96chromo}).  

If the source has an angular size that is large compared to that over 
which the magnification pattern varies, finite size (or extended source) 
effects will cause the subsequent light curve to deviate from the 
form given in Eqs.~\ref{feq}-\ref{ueq}.  In particular, the peak 
of the light curve will be somewhat flattened 
(\cite{wittmao94}, \cite{peng97}), an effect that has 
been reported in one case (\cite{alcockmb9530}).  In particular, 
whenever a point-lens or any part of a caustic curve transits the source 
face, the lensing structure becomes a large-aperture, high-resolution 
telescope, selectively magnifying some parts of the source much more 
than others.  Since the resultant light curve 
becomes broadened by an amount that depends on the size of the star 
(which can generally be determined through spectral typing) 
and on the transverse speed $v_\perp$, high-precision measurements of 
transit events can determine the relative proper motion of the lens, 
thus partially breaking the degeneracy otherwise present in Eq.~\ref{ueq}. 
Furthermore, since stars are limb-darkened by an amount 
that is wavelength dependent, 
color variations of a few percent 
can also be expected during transit events, and 
precise measurements of the resultant photometric and spectroscopic anomalies 
can lead to powerful constraints on stellar atmosphere models 
(\cite{loebsasse95}, \cite{gouldwelch96}, \cite{sassiap}, \cite{vallsgabaudchromo}).

\subsection{Non-rectilinear Motion: Parallax and Rotating Binaries}

Finally, a deviation from non-rectilinear motion in any one of the 
components that is significant over the duration of the microlensing 
signal will create an anomalous light curve. 
The orbit of the Earth around the Sun causes a so-called 
``parallax-shift'' in {\it every\/} lensing event that, if well characterized, 
can be used to determine the transverse velocity of the lens-source system in 
the ecliptic plane.  Such parallax events have been reported (\cite{alcock95parallax}), but reasonable assumptions lead to the conclusion 
that hourly sampling and $\sim$1\% photometry are required to 
achieve a 10\% detection rate (\cite{buchalter96para}).  
The measurement of parallax anomalies would provide crucial information 
in breaking the standard mass-distance-velocity degeneracy and 
thus determining the population from which Galactic lenses are drawn. 
Finally, non-rectilinear motion is expected within a binary source or binary lens system, and depending on the binary geometry may be detectable 
if the event is sufficiently monitored 
(\cite{hangould97}, \cite{pac97bin}, \cite{dom98rotbin}). 


\section{The PLANET Collaboration} \label{collab}

Although a few instances of most of the anomalies described 
in \S\ref{anomalies} have been observed, routine detection has 
been hampered by lack of continuous, high-precision, high-temporal 
resolution monitoring.  
PLANET, the Probing Lensing Anomalies NETwork, is a worldwide 
collaboration of astronomers whose primary goal is to provide this 
monitoring in order to detect and study microlensing anomalies, 
with the particular goal of determining the 
frequency and nature of lensing planetary 
systems in the Milky Way (\cite{albrow96}, \cite{albrow97}).  
PLANET was constituted in early 1995, soon after the  
first international microlensing meeting in Livermore at which 
all the major detection teams pledged their intention 
to provide public real-time alerts. 
PLANET uses semi-dedicated 1m-class telescopes at 
widely-separated longitudes in the southern hemisphere in 
coordinated monitoring campaigns.  The network 
is capable (in good weather) of providing nearly round-the-clock 
monitoring of several microlensing events a night; the detection 
and alert capabilities of the survey teams ensure that these events 
will be available.  

Keying on the electronic alerts provided by the MACHO and OGLE 
teams, PLANET is able to focus on individual events 
thereby achieving higher sampling rates and often higher precision 
photometry than do the detection surveys.   In order to optimize their 
detection sensitivities,  survey teams typically photometer 
several million stars nightly by imaging tens of fields in the Galactic 
Bulge.  Their exposure times are adjusted to the median stellar brightness 
and sampling times are dictated by the need to monitor as many fields as 
possible.   For the MACHO team, this has typically meant 
sampling intervals of $\sim$24 hours or more for Galactic Bulge sources.  
As discussed in \S\ref{results}, PLANET photometry is in contrast 
at least 10 times more frequent.  
The frequent, precise, multi-band photometry of PLANET is especially  
sensitive to the anomalies described in \S\ref{anomalies} that 
are short-lived compared to the duration of the primary event 
(source resolution, planetary anomalies, caustic crossings) and 
those that can present only small or subtle amplitude variations compared 
to the standard microlensing curve (planetary anomalies, blending, 
parallax, source resolution). 
In order to obtain stable, precise photometry with the shortest 
exposure times in all phases of the moon in these dusty Bulge fields, 
the Cousins I-band has been chosen as the primary PLANET 
monitoring band, but additional Johnson V-band monitoring is also performed 
in order to allow the detection of chromatic anomalies. 

PLANET has completed three Galactic Bulge observing seasons; 
results from the 1995 pilot campaign are presented here.   


\section{The 1995 PLANET Pilot Campaign} \label{campaign}

During the 1995 pilot campaign, PLANET had nearly continuous 
access for four weeks in June-July to four southern telescopes 
at three sites: the Dutch 0.91m and the Bochum 0.6m telescopes 
at the European Southern Observatory (ESO) on La~Silla, the South 
African Astronomical Observatory (SAAO) 1.0m at Sutherland, South Africa, 
and the Perth Observatory 0.6m at Bickley in Western Australia. 
Results from the Dutch 0.91m, SAAO 1m, and the Perth 0.6m are 
presented here; the observational parameters for these telescopes 
are summarized in Table \ref{telescopes}.   Beginning 
with the 1996 season, the Canopus 1m telescope near Hobart, Tasmania 
joined PLANET, greatly improving the longitude coverage of the network.

The large number of real-time microlensing alerts issued by 
MACHO (\cite{pratt96}) and OGLE (\cite{udalski94}) 
ensured that PLANET telescopes 
were continuously observing on-going microlensing events whenever the 
Bulge was visible (see Fig.~1).  
Dense monitoring can be performed only after alert; 
the post-alert portions of the 1995 curves falling 
within the 1995 PLANET season are shown as solid lines in Fig.~1. 
In total, PLANET monitored 11 events towards the Galactic Bulge 
in 1995, 10 during the primary campaign in June and July, 
and one during scattered 
observations in September, when a few baseline points were obtained 
for earlier events.  Light curves for 10 of these 11 events are 
presented here; since only a few data points were collected for 
MACHO 95-BLG-25 it has been excluded from this analysis.  

Mountain-top reduction proceeded in near real-time 
using DAOPHOT (\cite{daophot}) at La~Silla and Perth, 
and \dophot\ (\cite{dophot}) at 
SAAO.  All sites performed their own reduction and communicated 
with one another almost daily so that the observing strategy could 
be revised as necessary.  In order to track the progress of the  
events in all weather conditions and to facilitate inter-site communication, 
ten stars were chosen in each field as secondary standards.  
The flux of the microlensed source was expressed as a fraction of the 
average flux of these reference stars which was calibrated later 
against photometric standards.  The 
behavior of the reference stars also served as a guide to reduction 
difficulties with a particular image due to poor seeing, guiding errors or 
transparency fluctuations.

\vskip -0.4cm 
\epsfxsize=\hsize\epsffile{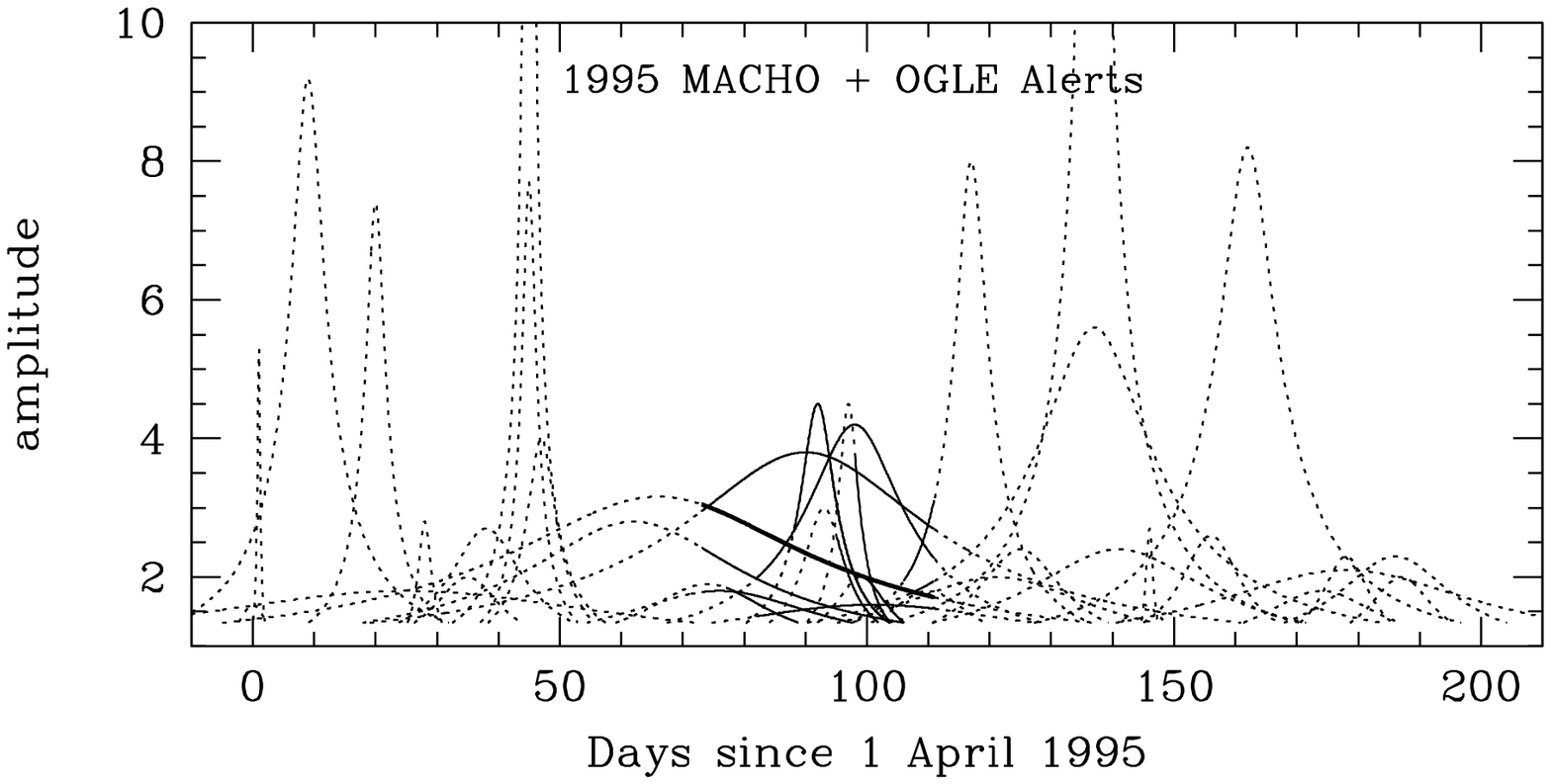}
\vglue -8.0cm
\figcaption{\tightenlines
Light curves (total magnification as a function of time) 
are shown for the 1995 real-time electronic alerts given by 
the MACHO and OGLE microlensing detection teams.  
Parameters for the light curves are 
taken from the MACHO alert page at http://darkstar.astro.washington.edu.  
The 1995 PLANET pilot season corresponded to days 73 to 111 on this plot, 
during which the majority of on-going events were monitored.  
}

Individual light curves were sampled 
every 1--2 hours; monitoring proceeded primarily in the 
I band, with occasional V observations interleaved. 
Exposure times were varied according to the conditions, phase of 
the moon, waveband and event magnitude, but were typically 5 minutes in 
I and about twice as long in V.    
A summary of the observations from each site is presented in Table 2.
In Fig.~2, a histogram of the interval between successive PLANET 
observations of a given event, summed over all events monitored in 1995,  
is shown for both of the primary observational bands, indicating that 
a median sampling time for each event of about once per 2 hours in 
the I band and once per 10 hours in the V band was realized over the 
duration of the campaign.  
Although the primary peak in the histogram of Fig.~2 illustrates 
that the longitude coverage of PLANET telescopes is such that 
$\sim$2 hour sampling was generally possible, 
the broad secondary peak near 18 hours is an indication that primarily 
poor weather, and to a lesser extent scheduling constraints (Table 1),  
occasionally limited sampling to one or two sites only.   

\hglue 1cm\epsfxsize=5.3in\epsffile{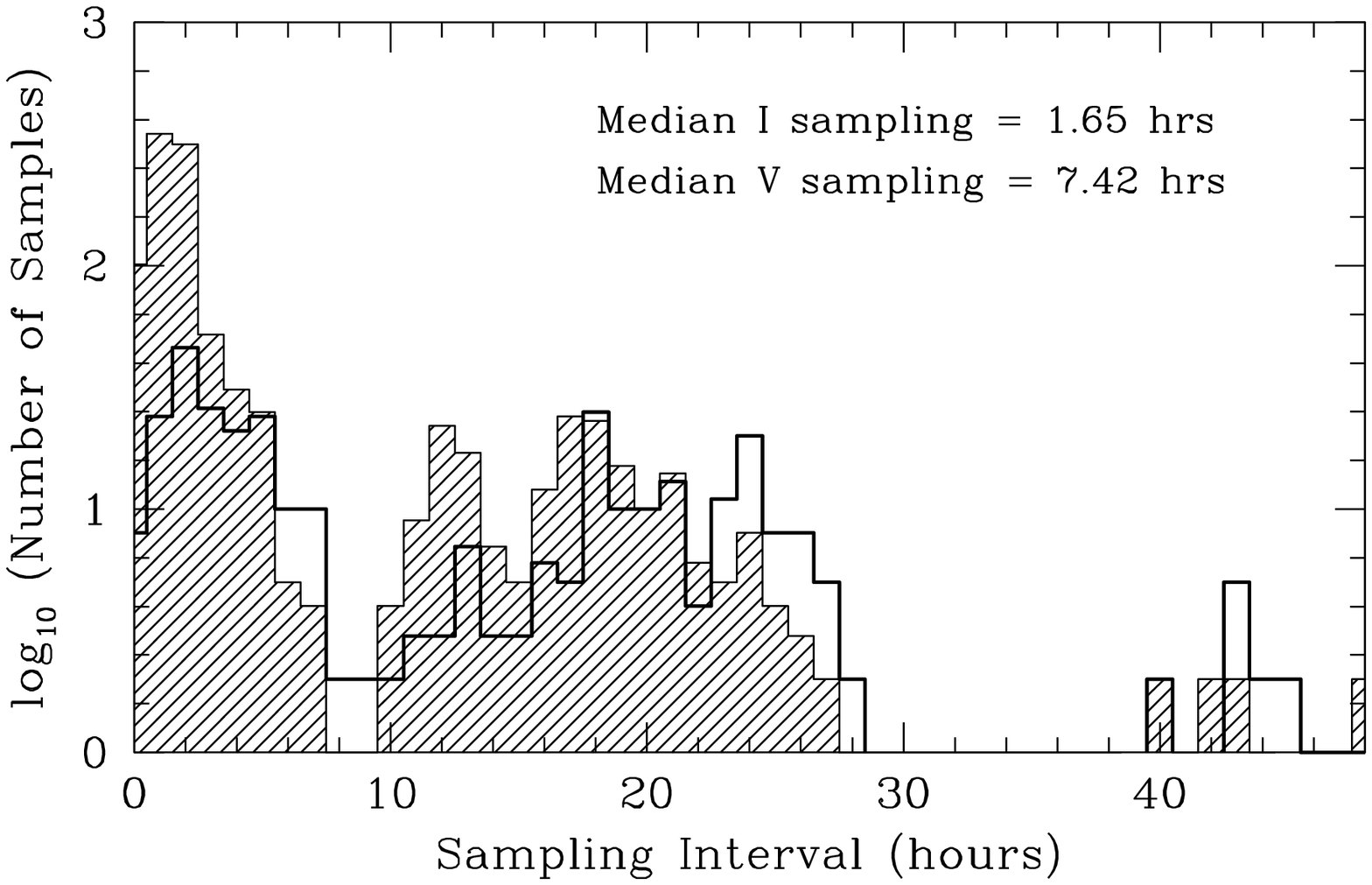}
\vskip -5.4cm 
\figcaption{\tightenlines 
Histogram on a logarithmic scale showing the time between PLANET 
photometric measurements in 1995, all three  
stations combined.  The shaded histogram shows the 
I-band sampling, which has a median value of about 1.65 hours; the 
open histogram with darker border shows the V-band sampling with 
median of about 7.42 hours.  
}

A description of the final data reduction with \dophot\ and a 
discussion of the resulting photometric precision achieved in these 
crowded fields are presented in \S\ref{reduction} and \S\ref{errors}.  


\section{Data Reduction} \label{reduction}

Since the prime aim of the campaign was to respond to anomalies in the light 
curves and to measure them precisely, it was important to be able to carry out
photometry at the telescope. As soon as images were obtained, they were 
de-biased and flatfielded with mean flats previously obtained at twilight, in
preparation for measurement. At La~Silla and Perth, 
the DAOPHOT (\cite{daophot})
package was used to photometer the frames, while \DOPHOT\ (\cite{dophot})  
was preferred at SAAO --- this difference of
approach reflected the exploratory nature of our first observing campaign; 
each site had had experience with a particular program, and it was not obvious 
at the outset which would be more suitable.  
In the end, DAOPHOT proved too 
slow to allow real-time reductions to be performed.  

Ten comparison stars near 
the lens were chosen by the La~Silla observer from the first frame obtained for
a given event. By referring the measured brightness of the lensed object to
the mean brightness of these relatively bright, uncrowded stars, the observer
produced over the course of the run a light curve that was 
independent of observing conditions, In addition, 
relative photometry from the different sites could be combined easily. 
This approach proved to be highly successful, and resulted in 
the discovery of the binary nature of MACHO~95-BLG-12 by PLANET while 
observations were in progress.

Since the many other stars on each frame apart from the lensed 
object can be used for variable star and Galactic structure studies, 
it was decided to reduce all the frames again in a more
consistent manner.  The crowded field photometry package \DOPHOT\ (Version~2) 
was used for this final data reduction. To improve photometric precision,
a catalog of positions was used for objects in each microlens field. This 
catalog was derived from a full reduction of the best quality image for the 
field. All other images were first partially reduced to give positions for the 
brighter stars; these positions were used to derive geometric transformations
with respect to the reference image. On any given image, objects were measured 
only at the transformed catalog positions. This reduces the number of \DOPHOT\  
fitting parameters from 7 to 5 for the brighter stars and to 2 for the fainter
ones.

To set the relative magnitude zero point, the ten reference stars per field 
chosen during the observing run were used.  Of these, 
only the ones that were relatively uncrowded (\ie\ \DOPHOT\ type = 1 in all 
but the very worst observing conditions) and stable were selected for the 
final reduction.  For nine fields, these criteria eliminated 
10 of the 90 original mountain-top 
reference stars; two were clearly variable and two others possibly so, 
while the remainder were judged to be too crowded. 
Due to its brightness, new (quite bright) 
reference stars were chosen for the MACHO~95-BLG-30 field.    

\subsection{Combination of Data}

Discussion of the photometric behavior of a given lens is best done in terms 
of the differential photometry described above, 
for which the data precision is highest. 
The different filters and detectors employed at each observing site 
could present potential problems when combining data from all sites. 
Differences between the effective wavelengths of the particular 
filter/detector combinations used at two sites would lead to
a color-dependent difference between their light curves for a given object.
Comparison of the SAAO and La Silla~data for the reference stars in all our 
fields shows that they are related by

$v_L = v_S - 0.086 \, (v-i)_S + C_v$
 
$i_L = i_S - 0.005 \, (v-i)_S + C_i$,

\noindent 
where lower case letters denote  magnitudes in the natural systems as 
observed at La~Silla (L) and SAAO (S), and $C_v$ and $C_i$ are constants. 
These equations imply that while the $i$ filters at the two sites 
had similar passbands, 
the $v$ filter at La~Silla had a somewhat redder effective 
wavelength that the one at SAAO.
Since the lens magnitudes were obtained 
differentially with respect to the mean magnitude of the reference stars,  
when combining data we have

$v_{L,lens} - < v >_{L,ref} = v_{S,lens} - < v >_{S,ref} - \, 
0.086 \, \left[ (v-i)_{S,lens} - < v-i >_{S,ref} \right] $

\noindent 
where $< \, >$ denotes the mean magnitude or color. 
A similar expression, with a coefficient 0.005, is obtained for $i$. 
Thus, there should be little effect from differences in filters 
and detectors for the $i$ magnitudes and, provided the mean color of the 
reference stars is sufficiently close to that of the lensed object, only 
a small effect for the $v$ magnitudes.

The discussion above concerns the relation between La~Silla and SAAO 
photometry only.  Sufficient data is not available to study the 
relationship between photometry at these sites and at Perth.  Perth 
contributed data for two microlensing events, for each of 
which a simple offset was sufficient to align the light curves 
photometrically.  The I-band filters of all three sites 
are thus probably quite similar.

\subsection{Transformation to Standard System}

To compare our data with those produced by other groups it is necessary to 
convert all data to a common, standard system, which requires transformation 
equations and an inevitable loss of precision.  Standardization is also 
necessary for astrophysical studies based on stars other than the lens
in our observed fields, for example the distribution of reddening, metallicity 
and stellar populations.  
Almost all of the standard star observations during the 1995 pilot 
campaign came from La~Silla, but since absolute photometric calibration 
had lower priority than monitoring many lenses with the highest 
possible time resolution, these data were insufficient for 
determining the color coefficients for the transformation equations 
for that site. 

At La Silla, all of our fields were referred directly to Landolt 
equatorial standards (Landolt 1983), while at SAAO 
only one field, namely MACHO 95-BLG-12, was standardized, in this case with 
respect to E-region standards (\cite{menzies89}). 
The transformation equations from the SAAO natural system to the
standard system are, however, 
known from observations made in other contexts to be

$v_S = V + 0.040 \, (V-I) + Z_{S,V}$ ~~~~{\rm and}

$i_S = I - 0.043 \, (V-I) + Z_{S,I}$.

\noindent 
Combining these equations with the ones above for the reference 
stars implies that the La~Silla transformation equations should be

$v_L = V - 0.053 \, (V-I) + Z_{L,V}$ ~~~~{\rm and}

$i_L = I - 0.048 \,(V-I) + Z_{L,I}$.

\noindent
Since the La~Silla data were insufficient to determine the 
color coefficients, the magnitudes reported in this paper were 
determined in effect from equations of the form 

$v_L = V + z_V$ and $i_L = I + z_I$,

\noindent 
in which case the zero points would depend on the mean
colors of the standard stars, ie,  

$z_V = -0.053 \, <(V-I)>_{standards} + Z_{L,V} = -0.053 + Z_{L,V}$ ~~~~{\rm and}

$z_I = -0.048 \, <(V-I)>_{standards} + Z_{L,I} = -0.048 + Z_{L,I}$.

\noindent 
In practice, the comparison between SAAO and La~Silla photometry 
yields the differences:

$v_S - v_L = 0.051 \pm 0.018 \, (s.d.)$ ~~~~{\rm and}

$i_S - i_L = 0.036 \pm 0.020 \, (s.d.)$,

\noindent
in good agreement with expectations.

The small standard deviations $(s.d.)$ imply that the reference stars in 
images of this field obtained at the two sites
were similarly affected by crowding in spite of the different focal plane
scales.  More work is needed to standardize PLANET data properly; 
the colors recorded in this paper are 
essentially on the natural system of La~Silla in the 1995 season.


\section{Photometric Errors} \label{errors}

Unlike most microlensing detection teams, PLANET does not adjust 
exposure times to  
mean conditions and mean field brightness in order to achieve reasonable 
photometry for the whole field.  Instead, in order to obtain the 
best photometry for the event itself, PLANET photometry is adjusted to the 
crowding of the event, its current brightness, and the observing conditions 
on a given night.  Consequently, PLANET photometric precision 
can be superior to that achieved by the detection teams, with 
relative photometry considerably more certain than the standardized 
photometry for most identified stars.  
Since the PLANET reference stars which were used as relative flux standards 
were chosen to be relatively bright (typically I$_C \sim$15 or 16) 
and uncrowded, the error in the relative photometry is dominated 
by the magnitude and 
crowding of the target star for all but the very brightest targets.  

\epsfxsize=6.25in\epsffile{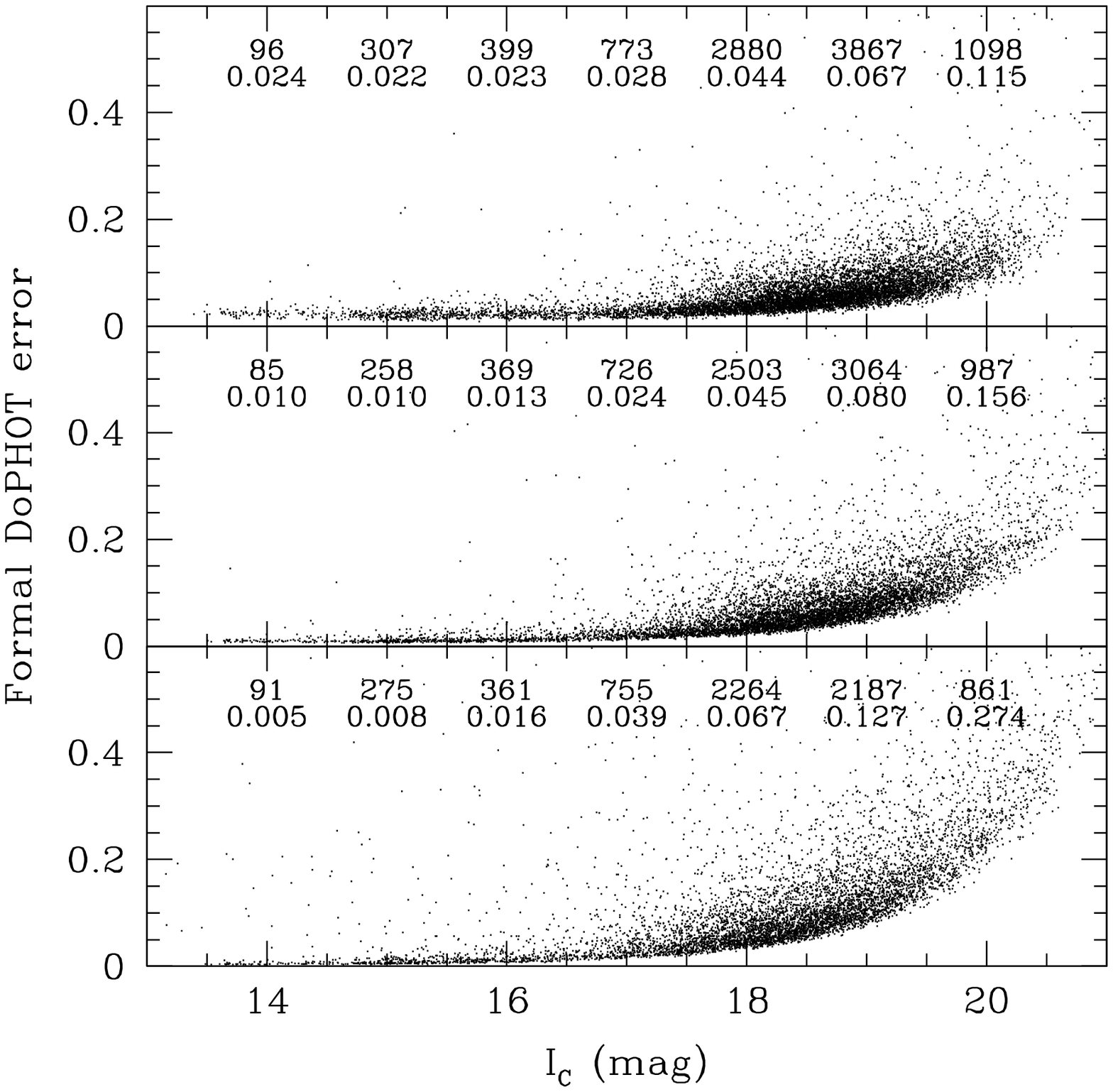}
\vskip -1.0cm
\figcaption{\tightenlines 
Formal \dophot\ error as a function of Cousins I magnitude (I$_C$) for La~Silla 
observations of the MACHO~95-BLG-12 field in three different 
seeing conditions of 1\asec1\ (top panel), 1\asec4\ (middle panel), 
which is typical for this site, and 2\asec2\ (bottom panel), representative 
of the worst seeing in which data was taken at La~Silla. 
The number of stars and median error (mags) in each magnitude bin 
is given above that bin.  
All measured stellar objects (\dophot\ types = 1, 3, 7) are plotted.
}

The fixed-position catalogs of \S\ref{reduction} were produced from  
the best-seeing frame for each field, which typically contained 
5000-10,000 stars, about 50-60\% 
of which could be measured reliably in typical seeing conditions 
at those sites. 
The densities of well-measured stars (\ie\ \dophot\ types 1, 3 or 7) 
in these catalogs were thus 
$\sim$0.03 ${\rm pixel^{-2}}$ or $\sim$0.25 ${\rm FWHM^{-2}}$, 
where FWHM is the full-width at half-maximum of the typical seeing disk.  
In these exceedingly dense conditions, crowding affects photometric 
precision.  Fainter stars, in addition to having smaller signal-to-noise 
ratios for a given exposure time, are typically also more crowded. 

Shown in Fig.~3 is the \dophot-reported error for the 
Dutch~0.91m (at La~Silla) field MACHO~95-BLG-12 as a function of the Cousins  
I-band magnitude (I$_C$) of the well-measured stars for three different 
seeing conditions.    
The exposure time and seeing of the frame displayed in the 
middle panel were typical for 
La~Silla at 5 minutes and 1\asec4, respectively.   
The top panel displays results for a frame in good seeing (1\asec1) and 
the bottom panel for a frame with quite poor seeing (2\asec2).   
In good seeing, uncertainties can be larger for bright stars,  
partly because exposure times are typically shorter, and partly because 
saturation of some pixels can result in a decrease in the signal-to-noise 
ratio.    
In the typical conditions, relative photometry with formal errors of  
$\sim$1\%, $\sim$2\% and $\sim$7\% were routinely 
obtained for I$_C$ = 15, 17 and 19, respectively.  
Poor seeing results in a smaller number of stars for which reliable 
measurements can be made, even when exposure times are increased.  
In addition, the reliability of photometry below 
I$_C \sim 18.5$ deteriorates rapidly.  
Comparison of the upper and lower panels of Fig.~3 indicates that 
photometry of the faintest stars that can be identified in these dense fields 
is limited more by crowding than by photon noise --- 
for stars with I$_C \sim 19$, magnitudes measured on the 2\asec2 seeing 
frame are about twice as uncertain as those from the 1\asec1 seeing frame, 
despite the longer exposure for the former.

Although the formal error determined by \dophot\ is 
meant to reflect the combination of uncertainties introduced by severe 
blending and photon noise, the actual scatter in a given constant 
star was generally greater than that reported by \dophot. 
As an indication of the reliability of the formal error, Fig.~4  
shows the ratio of the actual scatter over all frames for stars 
of a give I$_C$ to their mean formal \dophot\ error.   
All stars in the La~Silla field of MACHO~95-BLG-12 for which 
at least 100 measurements were available are included. 
The ratio is shown separately for those stars with well-behaved PSF 
(\dophot\ type = 1, top panel), ill-defined PSF (type = 7, middle), 
and nearby, partially-resolved neighbors (type = 3, bottom).  
Averaged over all frames, the true scatter appears to be $\ltorder$1.5 
times the mean \dophot\ error.  The ratio decreases with increasing 
magnitude as faint stars in the most crowded conditions (and thus with the 
largest and most-difficult-to-predict scatter) are no longer 
identified as point sources and fall out of the sample.  
In the mean, this ratio is not a 
strong function of \dophot\ type.

Systematic errors were also introduced by the phase and position 
of the moon, which especially affected the quality of the 
V photometry.   Nevertheless, useful data were 
obtained even in full moonlight conditions with the moon only 
$\sim$10\deg\ from the field center, in 
which case a severe gradient in the background was apparent over 
the field.  
High backgrounds from sources other than the moon, such as scattered 
light from clouds and reflections in the telescope optics also 
caused some deterioration of the final photometry, though this 
was seldom severe.  Small systematic effects correlated with 
air mass were also observed.  

\epsfxsize=\hsize\epsffile{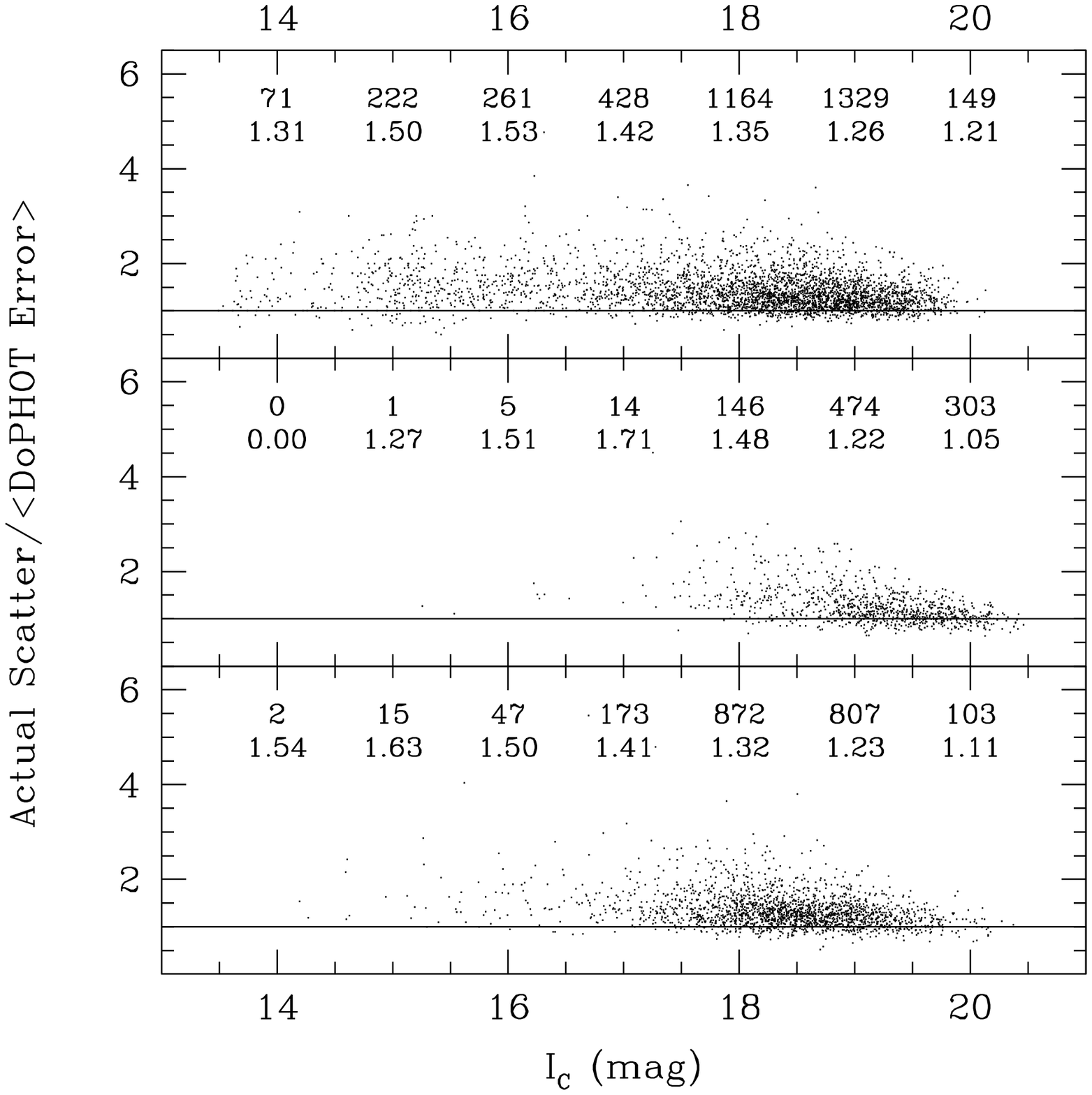}
\vskip -1.0cm
\figcaption{\tightenlines
The ratio of actual scatter to mean formal \dophot\ error 
is shown as a function of I$_C$ for the 6587 
stars in the La~Silla field of MACHO~95-BLG-12 for which 
at least 100 measurements were available. 
{\it Top:\/} Stars with well-behaved PSF (\dophot\ type = 1).  
{\it Middle:\/} Stars with ill-determined PSF (type = 7).  
{\it Bottom:\/} Stars with nearby, partially-resolved neighbors (type = 3).  
Numbers at the top of each panel indicate the number of stars 
and median ratio in each magnitude bin.
}

\vskip -0.35cm 
\epsfxsize=6.0in\epsffile{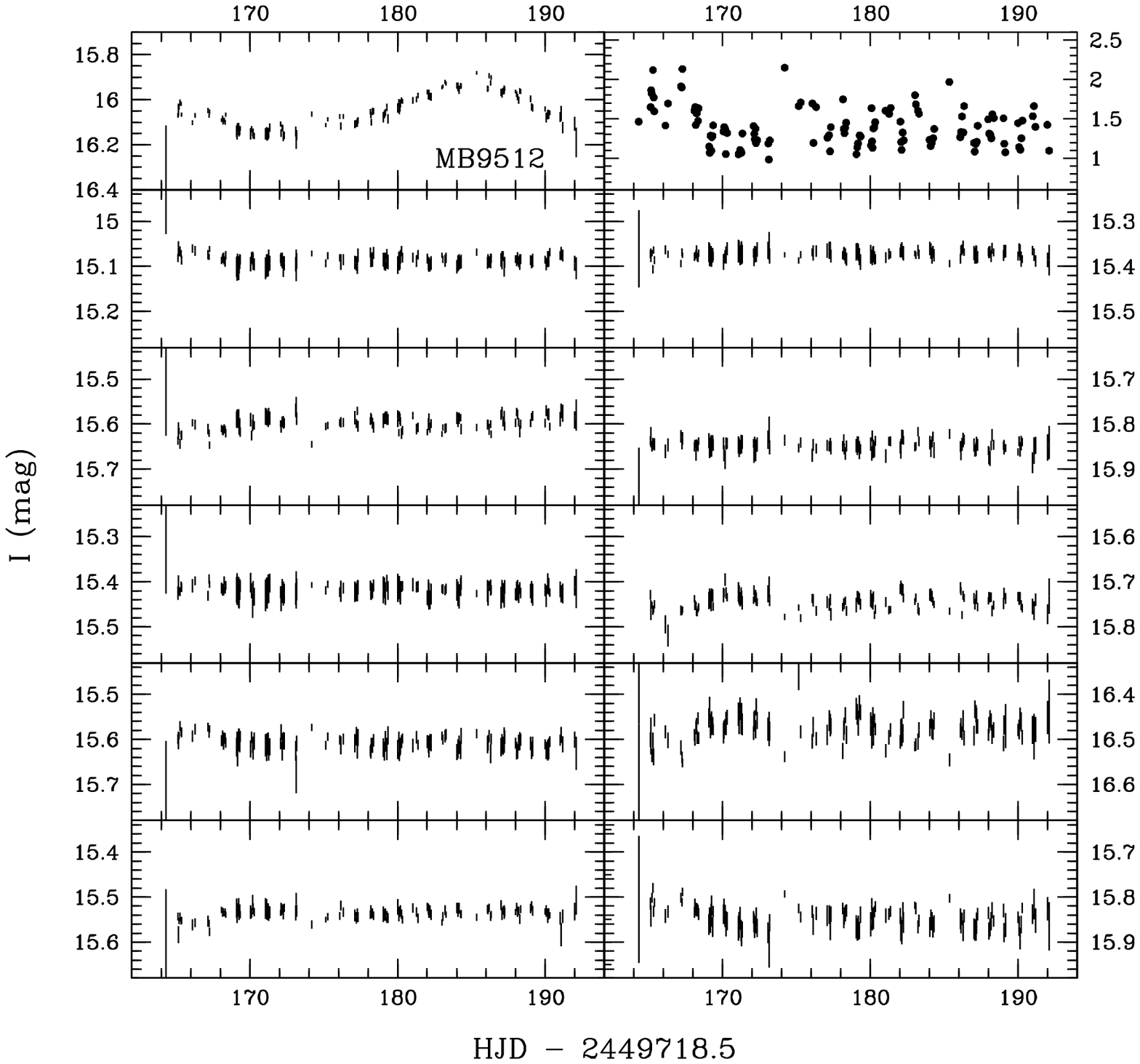}
\vskip -1.5cm 
\figcaption{\tightenlines
The La~Silla I band light curve for the binary event 
MACHO 95-BLG-12 (MB9512) is  
shown in the upper left-hand panel, together with the seeing for each 
frame in the upper right-hand panel and the light curves of 
the 10 reference stars in the lower panels.  Reference stars 
are displayed on a magnitude scale that is enlarged by a factor of 
two compared to that of the microlensing event.   
Formal \dophot\ errors are displayed.  Note that the apparently 
discrepant points in the event light curve near dates 174 and 185 
coincide with poor seeing, and are also discrepant in the 
photometry of some of the reference stars. 
}

Many systematic effects could be 
detected by examining the night-to-night behavior of constant 
stars, in particular the reference stars, in the same field.  
This technique was found to be so valuable for rejecting spurious 
systematic photometric deviations that PLANET on-line reduction 
automatically produces graphics like that displayed in Fig.~5 
to aid real-time analysis. 
The apparently discrepant points in the event light curve 
near dates 174 and 185 seen in Fig.~5, and the possible systematic effects 
seen near the beginning of this segment of the light curve 
coincide with poor seeing, and are also apparent in the 
photometry of some of the reference stars.   
Such systematic effects indicate the necessity of examining the 
light curves of constant stars in addition to that of the event 
(with associated computed error bars) when exploring the possibility 
that a microlensing anomaly has been detected.


\hglue 4.5cm \epsfysize=6.1in\epsffile{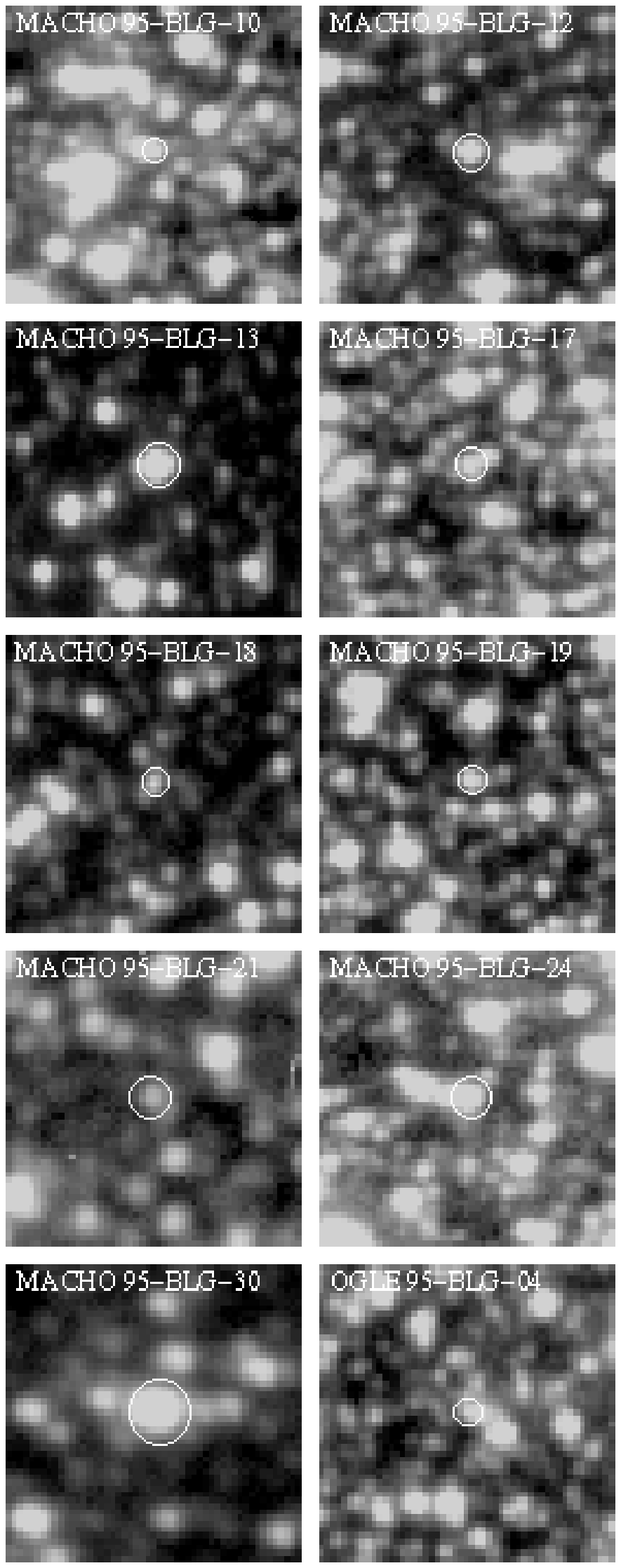}
\figcaption{\tightenlines 
PLANET finding charts for the microlensing fields 
monitored closely in the 1995 campaign; the source star 
is indicated with an open circle.  
The logarithmic contrast has been chosen  
to emphasize faint stars in the field and illustrate the crowding 
conditions for each event.  The MACHO 95-BLG-21 field is 23\arcsec\ 
on a side and was taken in 1\asec3 seeing.
All other images are 18\arcsec\ on a side and have seeing of 1\asec0 -- 1\asec1.
North is up; East to the right.    
}

\section{The Results} \label{results}

The rapid, precise, continuous multi-band monitoring required for our primary 
program is also suited to the discovery and study of many types of 
Galactic variable stars.  Searches for periodic light curves 
typically produce several variable stars per PLANET field, 
some with rather extreme or difficult-to-observe characteristics 
including periods of only a few hours, 
amplitudes under 5\%, or minima $I \sim 19$.  
The variable star studies will be presented elsewhere; 
here we focus on the results of our microlensing monitoring.

The 10 microlensing events closely monitored by PLANET in 1995 
and their immediate stellar environs 
are shown in the greyscale I-band images of Fig.~6.  
All frames were taken at the Dutch 0.91m at ESO with the exception of 
that for MACHO 95-BLG-21, which was taken at the SAAO~1m.  The contrast 
has been logarithmically adjusted to illustrate the crowding 
conditions of each source star.  Since the events 
were already in progress at the time of these PLANET observations, 
the microlensed star appears brighter (relative to its neighbors) 
than it would in a baseline finding chart.

Three of the 10 events (MACHO~95-BLG-10, 95-BLG-17, and OGLE~95-BLG-04) 
were so severely crowded as to appear to be optical 
doubles (\dophot\ type 3); in all seeing conditions their Point Spread 
Function (PSF) was too strongly blended with a neighbor to allow an independent 
photometric solution.  Fig.~6 illustrates this crowding visually, 
and light curves shown in 
Figs.~7 and 8 demonstrate that this is translated into larger 
photometric scatter than for 
unblended events (\dophot\ type 1) of comparable brightness.

\subsection{The Microlensing Light Curves}

The light curve of each monitored event is shown in Figs~7 and 8, 
in the I and V bands, respectively.\footnote{
All light curve data reported here, including those for the reference 
stars used in each field, can be found 
at the current PLANET WWW site: http://www.astro.rug.nl/$\sim$planet.}  
All data from all sites are shown except those for which the image quality 
was so poor that the event could not be associated with an identifiable 
stellar point spread function or one or more of the 
final reference stars was so severely blended so as to appear double.  
The event flux has been normalized in each frame 
by the average flux for the final reference stars chosen for that field.  
Calibration of the reference stars during photometric conditions 
then sets the absolute scale.   
Error bars reflecting the formal error returned by \dophot\ accompany 
each point in Figs.~7 and 8, and are often no larger than 
the size of the plotted point. MACHO~95-BLG-30 
was monitored outside the regular pilot season from La Silla 
only and thus has a different time scale, which is shown at the top of each 
figure.  

\vskip 1cm

\epsfxsize=\hsize\epsffile{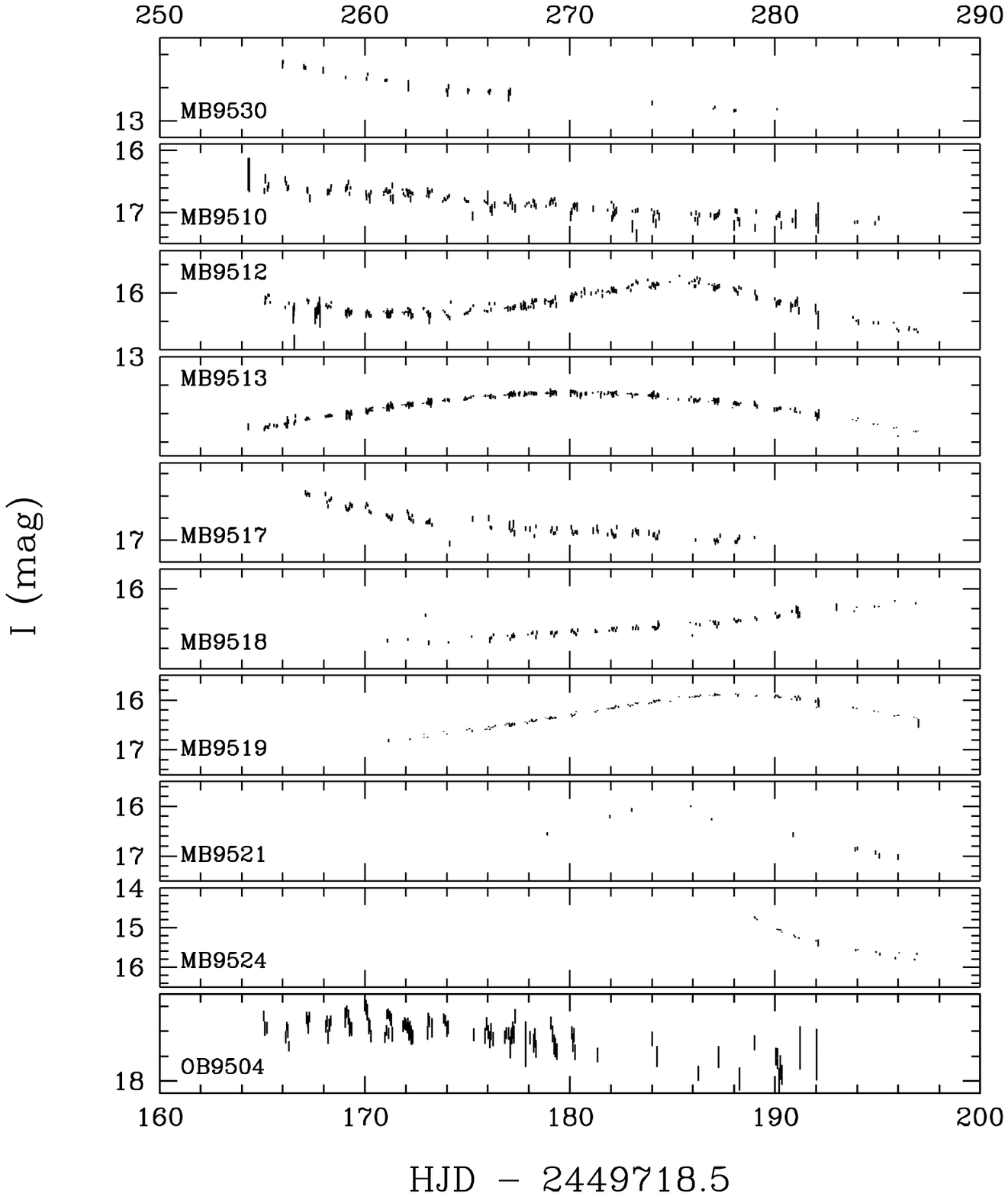}
\vskip -0.6cm
\figcaption{\tightenlines
Combined I-band light curves from all PLANET sites 
of the nine closely monitored events in the 1995 pilot season, 
and one extra event (MACHO 95-BLG-30) monitored by PLANET in 
additional observations later that year.   
The length of the error bar represents the formal \dophot\ error.  
All points are shown with no averaging or binning so that the 
scatter gives a correct indication of the true uncertainty 
in the relative photometry. 
For every plot, a small tickmark on the vertical axis represents 
0.2 magnitudes.  Abbreviated names for the events 
(\eg\ MB9530 = MACHO 95-BLG-30) are used to enhance legibility.
}

\epsfxsize=\hsize\epsffile{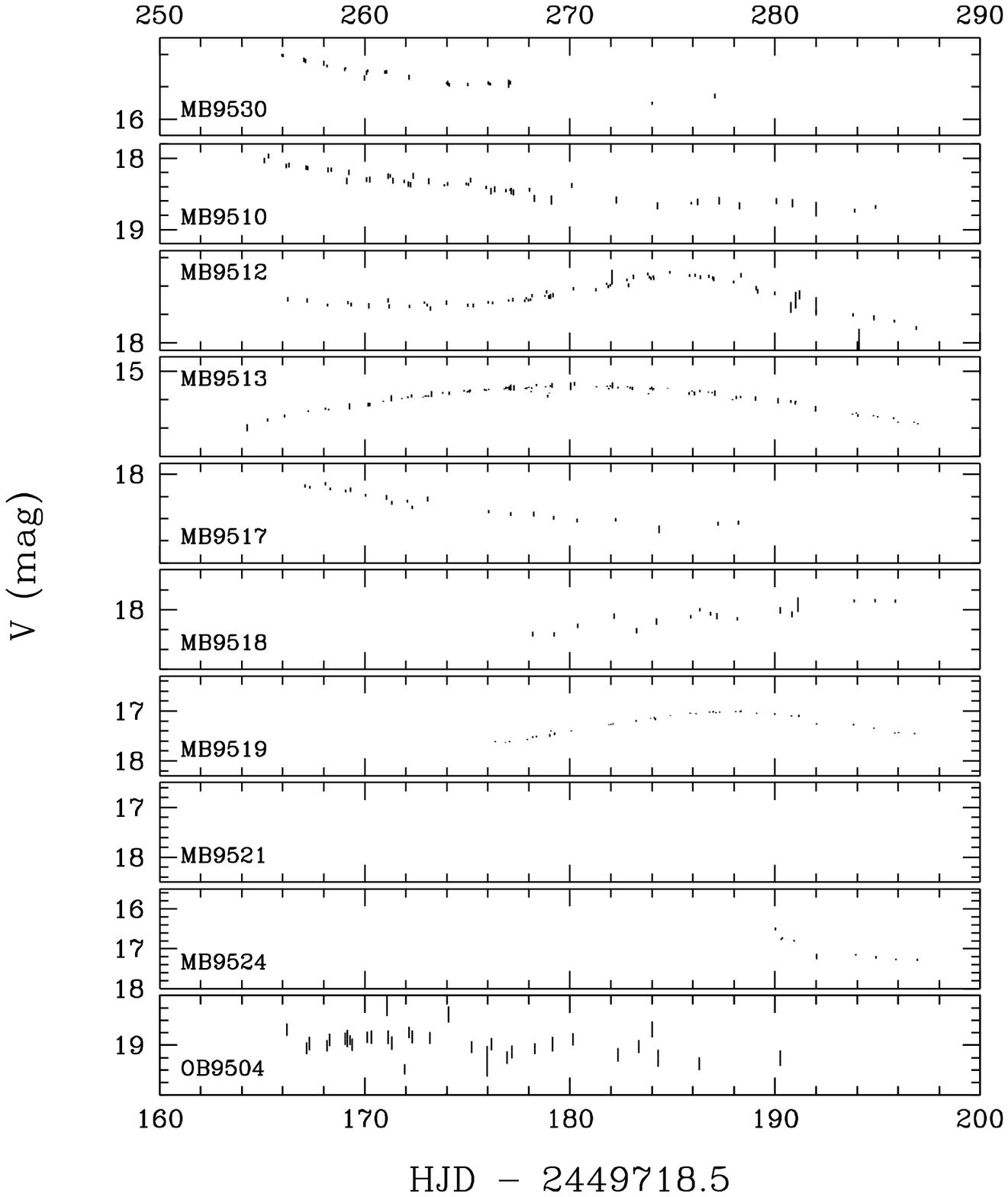} 
\vskip -0.6cm
\figcaption{\tightenlines
The same as Figure 5, but for V band.  Data from all sites are shown.  
For every plot, a small tickmark on the vertical axis represents 
0.2 magnitudes.  No V band data were obtained for MACHO 95-BLG-21.
} 

\vglue -0.25cm

\subsection{Fitted Event Parameters}

Relative photometry of stars in the monitored fields 
differed by up to a few percent among PLANET sites, 
even though the same set of reference stars was used.  
The offsets differed for different stars in the same field;  
no trend was discovered 
between the mean size or sense of the offset for a star 
and its stellar magnitude or color. 
Fainter stars exhibited larger scatter in their offsets, which 
may be because they are more likely to be severely crowded. 
Different detector resolutions and 
seeing conditions at each site coupled with the severe crowding in 
all of the fields may be responsible for these small site-dependent 
differences in 
relative photometry.

The size of the scatter in the offsets within a given field 
prevented the assignment 
of a fixed offset per field or per site. 
To determine the size of the offset for a given microlensing event, 
we further cleaned the light curve by removing all images in which the   
seeing was above 1\asec85 or the reference stars showed  
scatter much larger than the formal \dophot\ errors, 
which is an indication of unsatisfactory photometric reduction.  
This supercleaning procedure generally removed no more than 5-10\% 
of all reduced frames in each field. 
We then performed a combined fit to the supercleaned data from all 
sites within a given waveband with point-lens, point-source 
microlensing light curves (Eq.\ref{aeq}), allowing the   
site-to-site offsets (assumed to be multiplicative in flux) 
to float as free parameters for each band.  
(For MACHO 95-BLG-12, which shows clear indications of its binarity, 
the offset was determined by eye.) 
The incompleteness of our light curves caused by the 
finite length of our pilot campaign or post-peak alerts by the detection 
teams often resulted in ill-constrained fits to the standard point-lens,  
point source form, especially where data were missing at peak magnification 
or at baseline.  
Although a unique determination for all the standard microlensing parameters, 
$A_{max}$, $t_o$, $t_E$, and $F_o$, was generally not possible, 
the fitting procedure always produced robust, 
well-determined offsets, typically on the order of a few percent, 
even when different constraints were placed on the other lensing parameters.   
These offsets were used to photometrically align the multi-site light curve 
data shown in Figs.~7 and 8.

Although we were able to gather a few baseline points for most of 
our light curves in scattered observations after our pilot campaign, 
the relative baseline flux in the I and V bands from these 
observations is generally only known to within $\sim$10\%.  
The $V-I$ color of the source star in 
our own system of filters and reference stars can be determined 
quite accurately using the same method described above to 
determine offsets between multi-site data, but transformation 
of the color to the standard system 
introduces an additional uncertainty of about $\sim14$\%.  
Baseline magnitudes and colors for the events 
were determined from simultaneously fitting all 
supercleaned I and V data, and are reported in Table~\ref{planettable}. 
The magnitudes and colors are not de-reddened.  
Fitting uncertainties are indicated in parentheses.
Errors associated with transformation to the standard system 
have not been included as a more careful calibration 
of our fields is planned which will significantly reduce these uncertainties.
Only reasonably-constrained event parameters 
are listed in Table \ref{planettable}; in a few cases where 
our coverage was particularly insufficient, parameters 
that were poorly-determined by PLANET photometry were held fixed at 
their MACHO-derived values in order to constrain better 
the remaining parameters.   

\epsfxsize=\hsize\epsffile{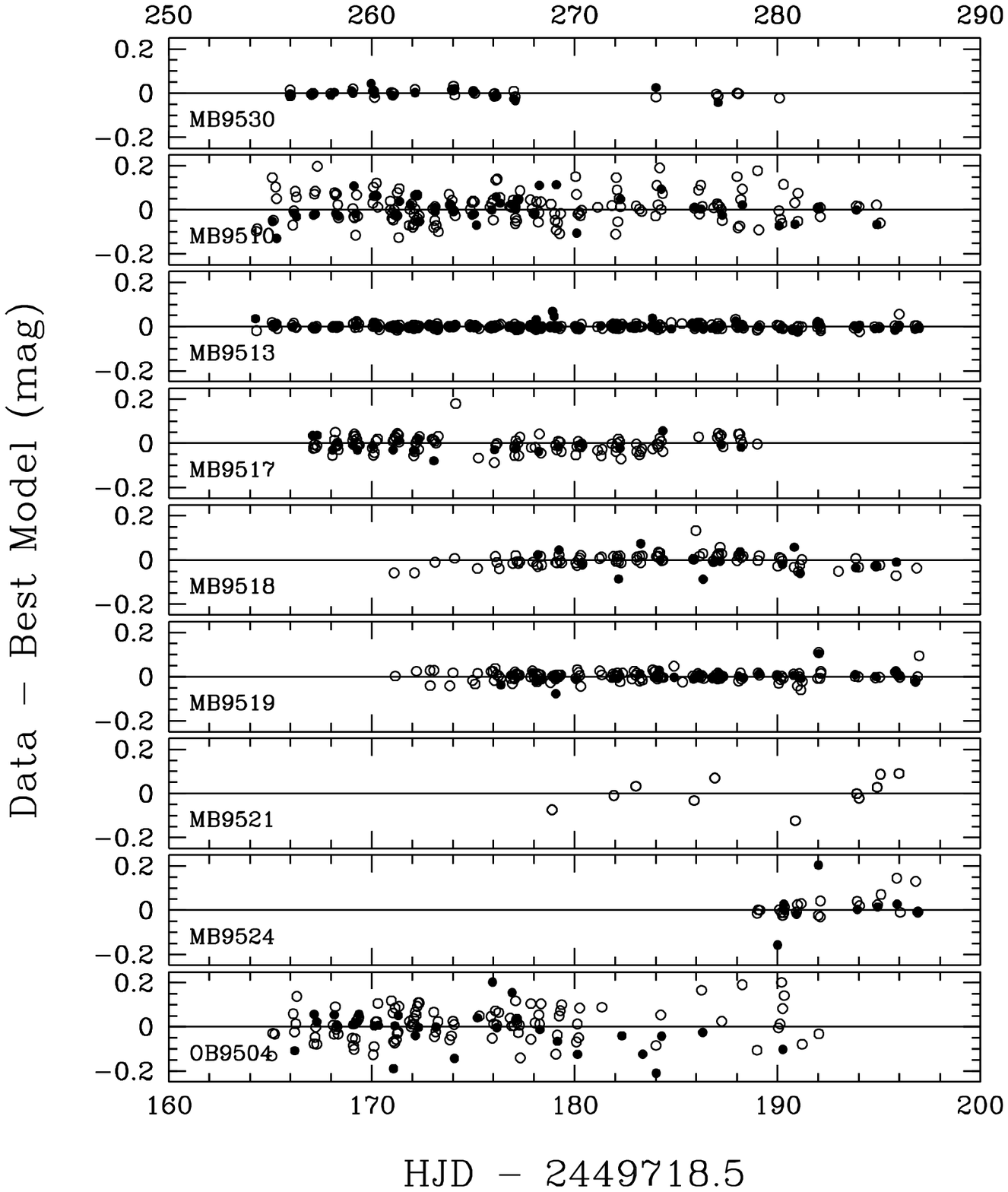} 
\vskip -0.5cm 
\figcaption{\tightenlines 
I- (open circles) and V-band (filled circles) 
residuals from the best point-lens 
point-source fit to all data are plotted separately to the scale 
for each event. 
All cleaned data are shown individually.  The scatter in the residuals 
for MACHO~95-BLG-13 and 95-BLG-19 is smaller than the size of the 
points, so that the hundreds of open circles indicating the I data are 
almost completely obliterated by the filled circles of the V data.   
The most crowded events clearly show the most scatter.  
} 

Residuals about these combined fits are shown in Fig.~9, 
where I (open circles) and V (filled circles) residuals are shown 
on the same scale for each event.  The 1-$\sigma$ scatter in 
the residuals ranged as small as 0.87\% for the I data of 
MACHO~95-BLG-13 to as large as 15\% for the V data of faint event 
OGLE~95-BLG-04, but were typically 4$\pm$3\% for the I data and 
5$\pm$4\% for the V data.  Note that this is consistent with 
expectations for the uncertainty in our relative photometry 
based on Figs.~3 and 4.

\vskip -0.75cm 
\hglue 0.25cm\epsfxsize=6.0in\epsffile{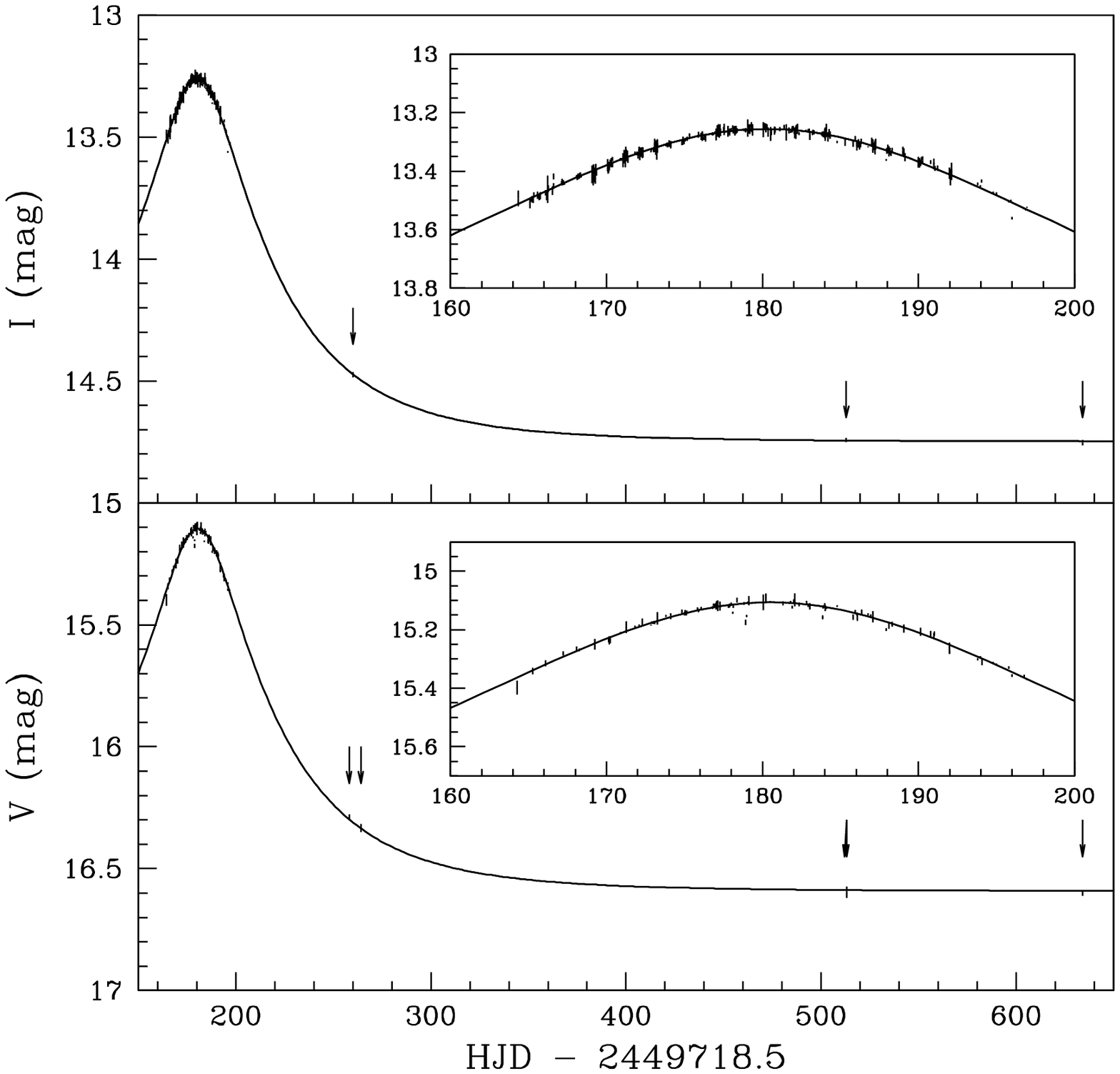} 
\vskip -1.2cm
\figcaption{\tightenlines 
Separate fits to the PLANET I and V-band data are shown 
for event MACHO~95-BLG-13.  All data are shown (without binning) 
with their associated formal \dophot\ errorbars.  Scattered baselines 
points are also indicated with arrows.  
Within the uncertainties, all event parameters are 
identical for the separate V-and I-band fits.}

The robustness of the fitted parameters for well-sampled events is shown in 
Fig.~10, which displays separate fits to the PLANET I and V-band 
light curves of MACHO 95-BLG-13 superposed on the data.  
No binning has been performed, so that the 
true scatter could be appreciated.  Errorbars are formal \dophot\ 
uncertainties.  An inset enlarges the region at peak magnification, and 
arrows indicate the positions of scattered baseline points 
obtained after the official pilot season whose small errorbars on this 
curve might go otherwise unnoticed.  The presence of these points 
near baseline significantly reduced the fit uncertainties.  
Within these uncertainties, all event parameters are 
identical for the separate V-and I-band fits. 

Comparison of Tables \ref{machotable} and \ref{planettable} indicate 
that the baseline magnitudes derived by PLANET and those derived 
by the detection teams agree within the uncertainties (including those from 
our transformation to a standard photometric system and the $\sim$10\% 
uncertainty in MACHO absolute photometry).  In addition, where we 
were able to perform relative photometry on MACHO (baseline) finding charts 
these also agreed with our fitted baselines to within the estimated 
uncertainties associated with transforming from MACHO~R to I$_C$. 

Without uncertainties for the values of $A_{max}$, $t_o$, and $t_E$ 
derived by the detection teams, it is difficult to assess the relationship 
between our fitted parameters and theirs, although there do appear to be 
a few real discrepancies.  In particular, the PLANET light curve 
for MACHO~95-BLG-18 in Figs.~7 and 8 is clearly rising beyond the 
MACHO-reported peak.  This may be related to the larger timescale 
and larger amplitude reported by PLANET for these events, if the MACHO 
estimates were not based on post-peak data.    
In addition, the event may be more blended in MACHO photometry than in 
PLANET images, resulting in decreased MACHO estimates for the 
timescale and amplitude.  
For two other events, MACHO~95-BLG-17 and 95-BLG-21, 
PLANET reports apparently shorter timescales than does MACHO. 
For this we have no explanation, but do note that we have three baseline 
points for MACHO~95-BLG-17, but none for 95-BLG-21.
The longer observing runs realized by PLANET in 1996 and 1997, 
and the additional baseline photometry taken for events monitored in 
those seasons are expected to lead to microlensing parameters 
that are considerably better constrained for 1996 and 1997 events.

Given the precision and density of the photometry obtained over 
finite portions of 10 light curves monitored by PLANET in its  
1995 pilot season, the remainder of this section will be devoted to 
the two types of anomalies to which this data set is most sensitive: 
chromaticity and binary lenses with small mass ratios, such as those due 
to binary star systems or massive extra-solar planets. 

\subsection{Chromaticity}

In order to test for chromaticity in the light curves, 
we have binned the I and V data in 24-hour periods and then 
computed the V$-$I color deviation in this period relative to the mean 
color of the event.  
The results are shown in Fig.~11.  
This test could not be performed on MACHO 95-BLG-21 since 
no V data were available.  
MACHO~95-BLG-24 shows a large single departure near date 192, apparently 
redder on this day than on days previous or following.  
Examination of the light curves shows that the single V-band 
data point during this 24-hour period was anomalously faint.  
This appears to be related to the exceedingly high background of 
that particular V frame, which significantly degraded the photometry of 
some of the reference stars as well. 

\vskip -0.2cm
\hglue 0.9cm\epsfxsize=5.0in\epsffile{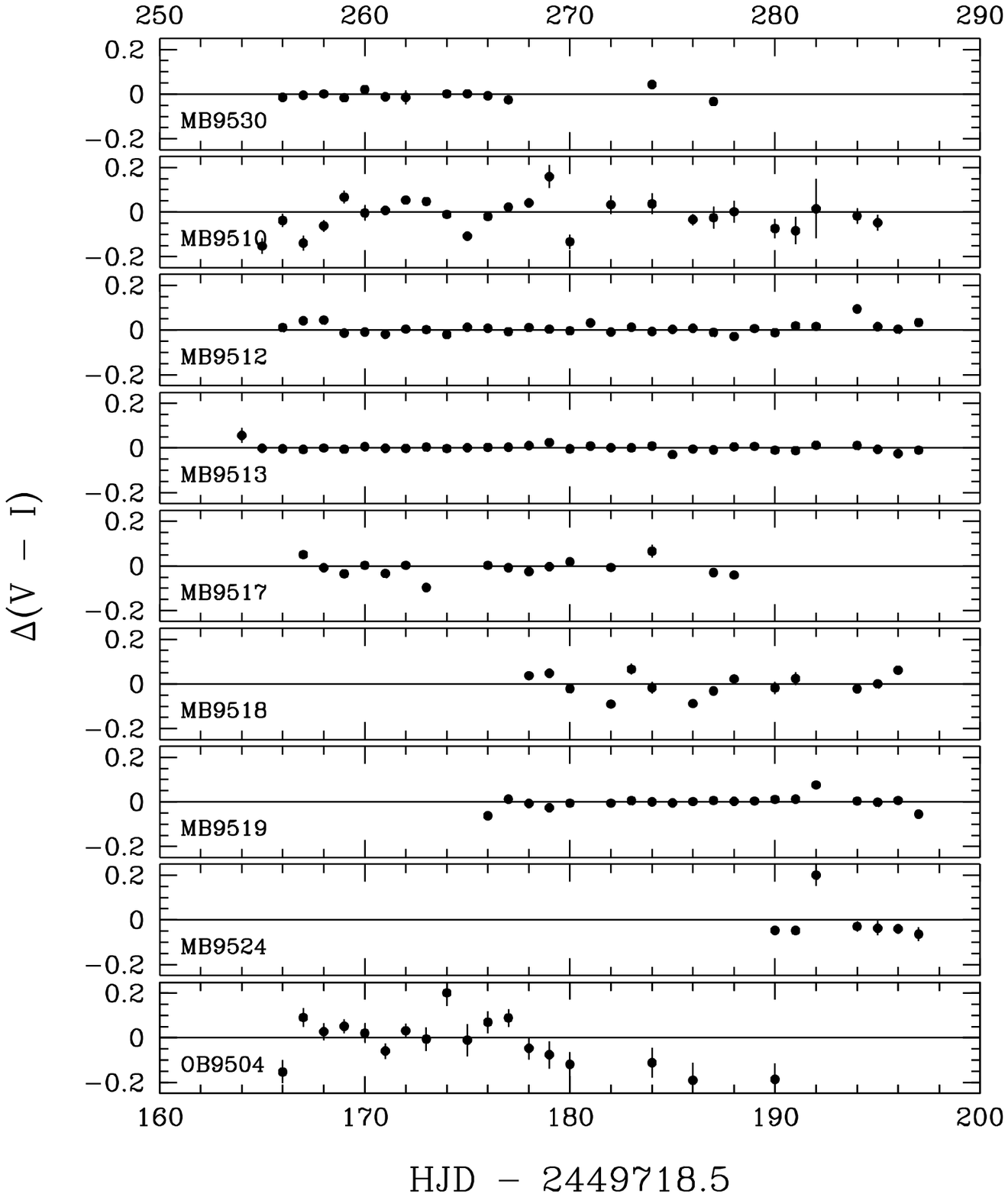}
\vskip -0.75cm
\figcaption{\tightenlines 
All cleaned I and V data have been binned into 24-hour segments 
and then subtracted to produce color deviations in magnitudes versus time.  
The mean color of the event is used as the reference color. 
No V-band data are available for MACHO 95-BLG-21.  The deviant 
point for MACHO 95-BLG-24 is due to high sky background in one V frame.  
OGLE 95-BLG-04 is 
the faintest and most crowded event in these data; this may result in 
the color shift with 
time as the microlensed star decreases in brightness while 
its companion(s) stays constant.
} 

The colors residuals of OGLE~95-BLG-04 in Fig.~11 show an apparent 
blueward trend with time, especially after date 178.  
Considering the ratio of the size of the discrepancy to the formal error 
in this and other events, the deviation on any given night could 
easily be spurious.  On the other hand, the color trend over the falling 
portion of the light curve for this, the faintest and most crowded 
of our events, may be an indication that the chromaticity is real, and due 
to blending with a near neighbor.  In that case, as the microlensed 
source decreases in brightness, its (presumably bluer) companion(s) 
remains constant, resulting in the observed chromaticity.  
The V$-$I color reported by OGLE for this event is 1.55, formally 
consistent with our value of 1.40$\, \pm\, $0.13, but also with the notion 
that the PLANET measurement is contaminated with light from blue 
companions.  This may also explain why the residuals 
are larger for OGLE~95-BLG-04 after date 180 in Fig.~9, and why 
the PLANET-reported I-band baseline (Table~\ref{planettable})  
is brighter than the OGLE value (Table~\ref{machotable}). 
Furthermore, fixing the baseline at the OGLE-derived estimate never resulted 
in satisfactory fits.  We conclude that OGLE~95-BLG-04 may be 
blended in some of these PLANET data.

With the exception of the single deviant point in MACHO~95-BLG-24
and the possible trend for OGLE~95-BLG-04, no clear trend of color with 
time is seen for the other events, with the 1-$\sigma$ scatters in 
the color ranging between 1.4\% for MACHO~ 95-BLG-13 to 7.2\% for 
95-BLG-10.  These are upper limits to the actual chromaticity for
these events since some artificial color deviation is 
introduced by the binning procedure, which does not account for 
the difference in the time during the night when the V and 
I frames were taken for these microlensing events.   
The I- and V-band residuals to the best-fit models presented in Fig.~9 
can be used to examine the chromaticity in a model-dependent, but 
point-by-point basis.

\subsection{Binarity of MACHO~95-BLG-12}

The binary nature of MACHO~95-BLG-12 is quite apparent and was 
discovered in real-time by PLANET during the course of its observations. 
Despite the strong departure of the light curve from that expected 
for a point source lensed by a point lens, this event shows 
no chromaticity above 2.6\%, averaging V$-$I in 24-hour bins 
(Fig.~11).  Nor does it show the gentle convexity that would be expected 
at one of the two peaks of a wide binary source event, but is instead 
more sharply peaked.   
This evidence is consistent with the hypothesis that MACHO~95-BLG-12 is 
a binary lens rather than a binary source, but does not constitute a proof.   
Combination of the densely-sampled PLANET data over the second peak 
and the more sparsely-sampled MACHO data over the entire curve 
(\cite{pratt96}) would be likely to result in a definitive model.

\subsection{Detection Sensitivity to Massive Extra-Solar Planets}

The discovery of the binary nature of MACHO~95-BLG-12 in the 
PLANET pilot season data makes it clear that PLANET monitoring 
is sensitive to the detection of binary lenses with 
components of similar mass.   We now discuss the detection  
sensitivity to binary lenses in which the ratio of the 
smaller mass $m_p$ to that of its primary lens $M$ is quite 
small, namely cases in which the partner has planetary mass.

As discussed in \S\ref{anomalies}, detection sensitivities to lenses 
with planetary systems have been estimated by a variety of authors 
who make simplifying assumptions about the nature and numbers 
of the planetary systems, the distance relative to the source 
and observer, the region of the light curve monitored, and the 
precision of the photometry.  Computation of the actual detection 
sensitivity of this PLANET dataset (or any other dataset) 
to planetary systems of various types is not trivial, and requires 
a computation, on a light curve by light curve basis, of the 
efficiency with which anomalies produced by a planetary binary of 
given mass ratio and geometry can be detected in the dataset. 
This efficiency will be a function of: sampling rate, photometric 
precision, maximum magnification of the event, 
placement and duration of the monitoring period with 
respect to the time of maximum magnification, degree of blending, 
and, for small mass ratios, the size of the source.
A full discussion of how these effects can be incorporated into 
realistic detection sensitivities is beyond the scope of this 
paper, but will be presented elsewhere (\cite{gaudisackett98}).
The estimate presented below is meant only to place the PLANET pilot 
season data in proper context by very roughly indicating its sensitivity 
to planets of given mass ratios; this also serves as an indication 
of what will be possible in the future. 

Not all light curves nor all portions of a given light curve 
are equally sensitive to the presence of a planet.   
Furthermore, the region of maximum {\it detection\/} will 
of course depend on the actual (unknown) orbital radii of the planets.  
Nevertheless within the Einstein ring ($A > 1.34$) of 
events of moderate magnification, typical of our dataset,   
light curve morphology is approximately  
equally sensitive to anomalies from so-called ``lensing zone'' planets, 
\ie\ those planets with projected separations from their parent star 
comparable to the primary's Einstein ring radius  
(\cite{gouldloeb92}).   
The ratio of the planet's mass $m_p$ to that of its primary, 
$q \equiv m_p/M$, affects both the duration of the anomaly and 
the frequency with which the anomaly will occur above a 
certain photometric threshold.   
The smaller the mass ratio $q$, the shorter the average duration and 
the less the chance of large photometric deviations in the light curve. 
The photometric precision of the data will determine the threshold 
above which anomalies can be detected.  
As long as the source is not resolved, the average duration scales 
with $\sqrt{q}$, and thus 
linearly with the Einstein radius crossing time $t_E$ of 
the primary lens.   The photometric sampling interval 
is thus naturally expressed in units of $t_E$ when discussing 
the sensitivity to planetary systems of a given mass ratio.  
If this sampling interval is sufficiently small for a given mass 
ratio, then the sensitivity to planet detection depends on the total 
length of light curve (in units of $t_E$) 
monitored with a given photometric precision 
(assuming that all portions of the monitored light curve are 
equally likely to contain an anomaly).   
 
\vskip -0.2cm
\hglue 1cm\epsfxsize=5in\epsffile{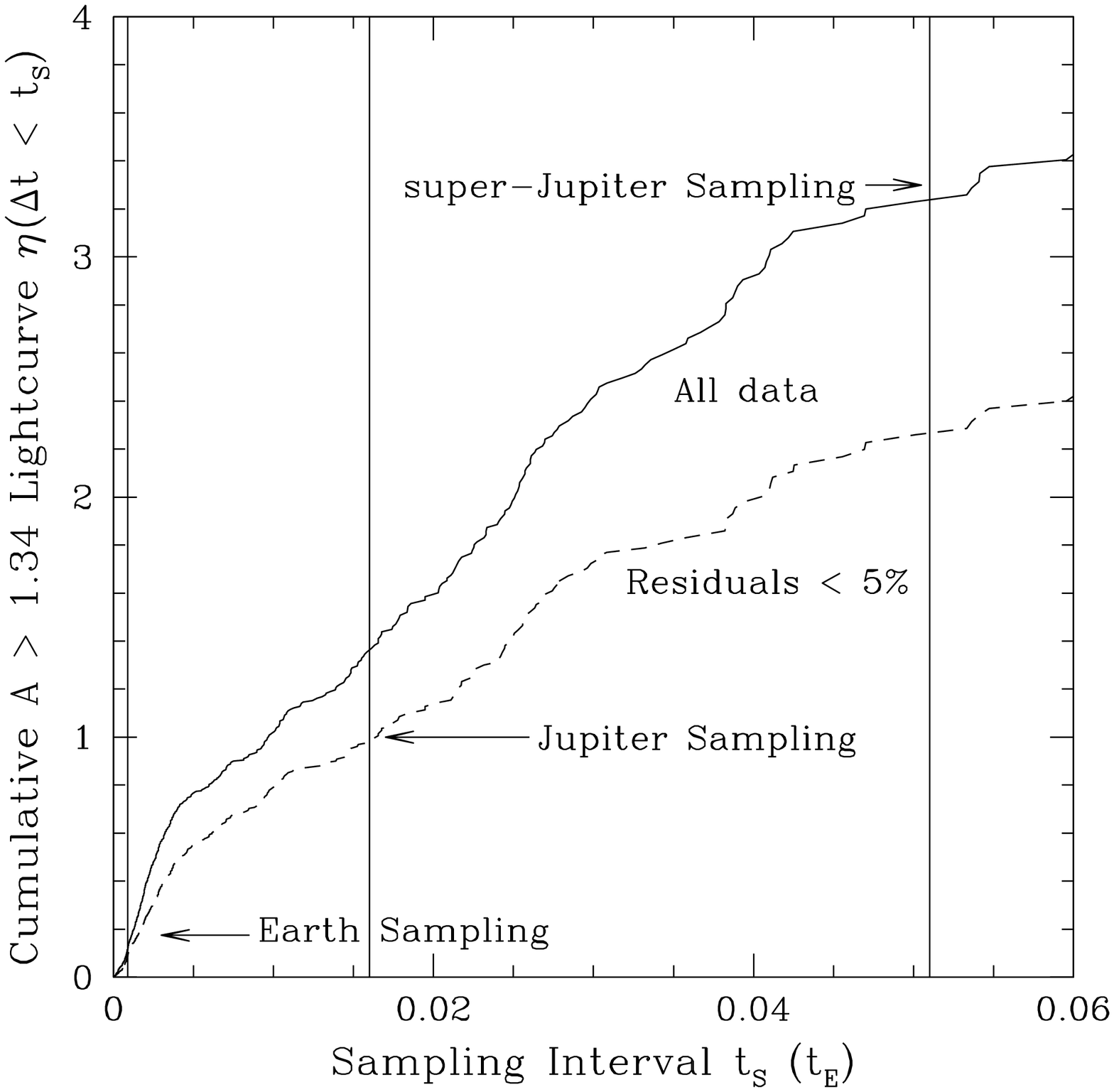} 
\vskip -0.5cm
\figcaption{\tightenlines 
Total length of normalized light curve $\eta$ 
sampled by PLANET during its pilot season as a function of 
the normalized sampling interval $t_S$ (in units of $t_E$).  The full 
curve shows all data; the dashed curve only those datasets that 
have true scatter of 5\% or less from the best-fit models.  
The vertical lines, which are explained more fully \S\ref{results}, 
give approximate 
minimum sampling intervals appropriate to detecting planets of a 
given mass ratio $q$. 
Light curve data of cumulative length $\eta \gtorder 6$ must 
be sampled in order to place constraints on the 
numbers of Jupiter-mass planets ($q \sim 0.001$); more massive  
planets can be constrained more easily, while lower mass planets 
require more cumulative light curve length.
} 

With these considerations in mind, 
we wish to combine the light curves for the events monitored by 
PLANET in the 1995 pilot season into one rough 
figure-of-merit for planet detection.  
To do so we have normalized the monitoring period of each 
light curve by $2 \, \sqrt{1 - u^2_{min}} \, t_E$ in 
order to obtain the fractional length of monitored curve 
inside the Einstein ring radius ($A > 1.34$). 
We then sum over all light curves at a given value of 
the normalized sampling time $t_S = \Delta t/t_E$.  
The cumulative normalized light curve length $\eta$ sampled 
at normalized intervals less than $t_S$ 
is shown as the solid curve in Fig.~12 as a function of $t_S$.  
The dashed curve is the same 
cumulative normalized light curve length for those datasets 
in which the 1-$\sigma$ scatter from the best-fit model 
is less than 5\% (namely, the PLANET I light curves 
for MACHO~95-BLG-12, 13, 17, 18, 19, 24 and 30, 
and the PLANET V light curves for MACHO~95-BLG-12, 13, 19 and 30).

The typical duration of a planetary 
anomaly is given very roughly by the time to cross the diameter 
of the {\it planetary\/} Einstein ring, $ 2 t_p =  2 \, \sqrt{q} \, t_E $.
Assuming a typical $t_E$ of 20 days, and a lensing system half way 
between observer and source, 
$ 2 t_p  
\approx \, 1.7 \, (m_p/M_{\earth})^{1/2} \, (M/\msolar)^{-1/2} \, {\rm hrs}
\approx \, 0.0035 \, (m_p/M_{\earth})^{1/2} \, (M/\msolar)^{-1/2} \, t_E $.  
In order to expect an average of four deviant points over the anomaly, 
the sampling interval should be no larger than one-fourth of this duration, 
or 0.0009 $t_E$, 0.016 $t_E$, and 0.051 $t_E$ for Earth-mass, 
Jupiter-mass, and 10 Jupiter-mass (super-Jupiters) planets orbiting  
solar-type stars, respectively.  These maximum sampling intervals  
are indicated in Fig.~12 as vertical lines.  
Assuming that the primary lens is a solar-type star half way to 
the Bulge, Fig.~12 shows that a total normalized light curve length 
$\eta \sim 3$ has been monitored with sufficient density to 
detect super-Jupiters, 
$\eta \sim 1$ has been monitored often enough to detect Jupiters, 
and almost none of the total light curve length has been monitored 
densely enough to detect Earth-mass anomalies.

If all the lenses monitored by PLANET in 1995 have planets with 
mass ratio $q = 1 \times 10^{-3}$, as would a Jupiter orbiting a
solar-mass star, would any have been detected? 
Gould \& Loeb (1992) have shown that, averaged over the portion 
of the light curve within one Einstein ring radius of the peak, 
the detection sensitivity to a Jupiter-mass planet at a projected 
distance of 5~AU from its 1~$\msolar$ lensing primary  
is $\sim$17\% if the system is located half way to the Galactic Bulge. 
This sensitivity assumes that deviations of 5\% can be reliably detected, 
in which case a cumulative normalized light curve length of $\eta \sim 6$ 
must be monitored to expect a single detection, even if all lenses have 
such planets.  Since we have adjusted the maximum sampling interval 
to expect $\gtorder \, $4 deviant points, 
it is reasonable to assume that the light curves with 5\% photometry 
(dashed $\eta$ curve in Fig.~12) should be sensitive to 
$\gtorder $5\% deviations at the 2$\sigma$ level.  
Even so, this simple 
estimate shows that the total length of $A > 1.34$ light curve 
monitored in the pilot season was probably 
insufficient by about a factor of six for Jupiter-mass sensitivity. 

On the other hand, these pilot season data may well be sufficient in 
monitoring frequency, total light curve length, and photometric 
precision to begin to place limits on the number of super-Jupiters 
orbiting Galactic lenses.  Such systems would have larger detection 
probabilities and thus require a total light curve length of 
$\eta \gtorder 2$ for reasonable detection probabilities 
(if all lenses have super-Jupiters in their lensing zones).  The 1995 pilot 
season data, especially in combination with the 1996 and 1997 PLANET data, 
should be suited to constraining the numbers of such massive extra-solar 
planets orbiting suns too distant to be seen directly.   
The existence of such planetary systems can be deduced only with 
the aid of microlensing monitoring. 
We stress that the discussion in this section is only intended 
as an order-of-magnitude estimate;     
more quantitative results await a full calculation of planetary  
models and observational efficiencies. 


\section{Conclusions and Future Outlook} \label{conclude}

The results of the 1995 PLANET pilot campaign conclusively demonstrate the 
feasibility, challenges and rewards of microlensing monitoring.
PLANET telescopes were continuously and 
fruitfully employed throughout the month-long 
campaign performing photometry with a precision and frequency 
not feasible for the current microlensing detection programs 
designed to provide the largest number of alerts.    

Despite overhead and weather, monitoring of several events 
simultaneously proceeded with median sampling times of about 1.6 hours 
in the I-band; 
V-band monitoring was about 5 times less frequent by design.   
Real-time mountain-top reduction to track event progress was successfully 
demonstrated, and by using reference stars in the field as relative 
flux standards, relative photometry was performed in a wide 
variety of weather conditions and the data successfully combined 
between sites on a daily basis. 
Examination of the photometry of these (constant) reference stars proved 
crucial in separating systematic effects in the photometry from 
real anomalous behavior. 
PLANET relative photometry was able to reach I=19 with an 
actual precision of 0.10mag and was 
ultimately limited more by stellar crowding 
than by photon noise.  

Two of the 10 events  
showed significant anomalous departures from the point-lens, 
point-source light curve.  The binary nature of MACHO~95-BLG-12 
was detected by PLANET in real time on the mountain.  Although 
not conclusive, the shape of the light curve near the second 
of its two peaks and the lack of V$-$I chromaticity over the 
PLANET-monitored portion of the light curve at the 2.5\% level 
suggest that this event is a binary lens rather than a binary 
source.  Definitive modeling will require combining MACHO and PLANET 
data for this event.  OGLE~95-BLG-04, the faintest and most 
crowded of the events monitored by PLANET, displayed chromaticity 
at the 20\% level over the last half of its PLANET-monitored light curve.  
This, together with the differences between OGLE and PLANET 
determinations of the fitted baseline and event duration parameters, 
suggests that this event may be blended in the PLANET data.  
The events showed scatter about 
their best fit point-source, point-lens light curve 
ranging from $<$1\% (for PLANET I-band photometry of MACHO~95-BLG-13) 
to about 15\% (for PLANET V-band photometry of OGLE 95-BLG-04), with most 
residuals at the 4-5\% level.  

Detailed calculations integrating 
taking into account the actual sampling times and photometric 
precision on a light curve by light curve basis will be required before firm 
estimates can be placed on the planet detection sensitivity of 
PLANET observations.  Nevertheless, simple considerations 
indicate that the total length of microlensing light curve sampled 
by PLANET in its 1995 pilot season falls somewhat short of 
placing constraints on the number of lenses with planets 
of mass 0.001 times that of their primaries (Jupiters), but may be 
sufficient to constrain the number of planets with mass ratios 
of 0.01 and higher (super-Jupiters).  The detection of Earth-mass 
planets would require very rapidly-sampled (sub-hourly), 
high-precision (1-2\%) photometry of hundreds of non-giant stars 
to place meaningful constraints.  

Looking toward the future, one can use the qualitative and 
quantitative lessons of the 1995 PLANET pilot campaign to 
plan more effective observing strategies and to give an indication 
of what might be expected in the coming years from intense 
microlensing monitoring.  

The number of events that can be 
monitored per night with telescopes of a given aperture depends on 
observing overhead, the brightness and crowding of the source star, 
and on the desired level of photometric precision.
To achieve the precision realized in the 1995 pilot season, 
PLANET telescopes monitored about 10 events of moderate brightness 
with better than 2-hour sampling over the period of one month.  
Since the typical duration of an Galactic bulge event 
(defined as $2 \, t_E$) is on the order of a month, the overall 
sensitivity to most microlensing anomalies will scale roughly as the number 
of months of usable monitoring observations.  
In subsequent 1996 --- 1998 observing seasons, 
PLANET has been granted increasingly 
larger blocs of observing time, making more of the Galactic bulge 
season accessible.  This has resulted in an approximate doubling 
of the number of events monitored in 1997, and a further increase 
of a factor of 1.5 to 2 is expected in the 1998 season.  
Longitudinal coverage has been significantly improved with the 
addition of the Canopus~1m telescope near Hobart, Tasmania to the complement  
of PLANET telescopes; Canopus also serves as a hedge against bad weather 
at the Perth longitude.  Finally, detectors of increased sensitivity 
have or are being installed at all PLANET sites.  Taken together, these 
enhancements will serve to increase the sensitivity to planet detection 
(per bulge season) substantially compared to the month-long 
1995 pilot campaign. 

Changes in monitoring strategy may also lead to increased sensitivities 
to planetary anomalies.  In the 1995 pilot season, sampling 
frequencies were nearly constant at once every 1 to 2 hours, as weather 
allowed.  In subsequent seasons, PLANET has made the conscious 
decision to scale the interval between adjacent measurements on each light 
curve with the Einstein crossing time $t_E$ for that event 
so as to increase the probability of detecting putative Jupiter-mass 
planets in the lensing zone of the primary.  In this 
way, the complete dataset should have a more uniform detection 
efficiency in this portion of phase space.  
Similarly, in order to spend most of the observing resource on the 
portion of the light curve that is most sensitive to planets in the lensing 
zone of a few AU, light curves are monitored intensely only while the 
source is projected within 1---1.5 $\theta_E$.  
Eventually, as the number statistics of intensely-monitored 
events increase, it may be sensible to relax these constraints in order 
to increase sensitivity to planets outside the lensing zone or 
planetary systems with extreme mass ratios. 

Lastly, event selection can play an important role 
in determining the success of achieving particular program goals. 
In all cases, photometric precision is important, so that whenever 
possible the choice of (relatively) uncrowded source stars is 
advisable.  Events should be chosen to have timescales $t_E$ short enough 
to fit comfortably into the observing season, but have post-alert 
durations long enough to allow sufficient sampling at the desired rate.
Low-impact parameter (high-magnification) events are also favored 
for planet detection 
since the source is always brought near the central caustic, 
increasing the chance of an anomaly.  
If the goal is to maximize the detection sensitivity to Earth-mass 
planets with mass ratios $q \sim 1 \times 10^{-5}$ to $1 \times 10^{-6}$, 
source resolution effects will dilute the amplitude of anomalies if 
the angular radius of the source star is larger than that of the 
planetary Einstein ring radius (\cite{bennettrhie96}); this selects against 
giants or sub-giants as source stars.  If, on the other hand, the 
goal is to maximize sensitivity to larger mass-ratio planets, 
giants and sub-giants are favored since their relative brightness 
will allow more precise photometry with shorter exposures without 
introducing a bias against lenses of a particular type.   
  
Given the encouraging results of its 1995 pilot campaign, the future of 
PLANET looks bright as it continues  
to monitor microlensing events in search of anomalous behavior.  
Each Galactic bulge season brings an increase in the numbers 
of electronic alerts being issued by the microlensing 
survey teams, allowing PLANET to be more selective in choosing 
events to match its program goals. 
Together with the increasingly generous awards of 
telescope time from PLANET-participating sites, this has resulted in 
an ever-growing database of precise and rapidly-monitored microlensing 
events, increasing the detection sensitivity to anomalies of all kinds, 
in particular those due to extra-solar planets in the inner Galaxy.


\acknowledgments

PLANET thanks the MACHO and OGLE collaborations for providing real-time 
electronic alerts of microlensing events in progress; 
copious early alerts are 
crucial to the success of our intensive microlensing monitoring.  
In addition, we would like to thank Abi Saha and Mario Mateo for 
their help in setting up \dophot.  
Financial support from ASTRON, the astronomical division of 
the Foundation for Nederlands Wetenschapelijk Onderzoek 
(Dutch Scientific Research), NWO is gratefully acknowledged.  
PLANET members also wish to thank The Leids Sterrewacht Foundation, 
Perth Observatory, and the 
South African Astronomical Observatory for the generous allocations 
of time that made this pilot program possible.




\vglue -1cm 
\begin{deluxetable}{lrrrrcc} 
\tablenum{1} \label{telescopes}
\tablecaption{
Observational parameters for the PLANET 1995 pilot observing season}
\tablehead{
\colhead{Telescope} & \colhead{Long} & \colhead{Lat} & 
\colhead{Pix Scale} & \colhead{CCD Format} & \colhead{Dates of Obs} & 
\colhead{Seeing\tablenotemark{a}} 
}
\startdata
Dutch 0.91m & 289.27\deg\ & $-$29.25\deg\  & 0.44\arcsec\ & 512 $\times$ 512 & 12 Jun -- 13 Jul\tablenotemark{b} & 1.40\arcsec \nl
SAAO 1.0m & 20.81\deg\ & $-$32.38\deg\  & 0.35\arcsec\ & 512 $\times$ 512 & 20 Jun -- 17 Jul & 1.65\arcsec \nl
Perth 0.6m & 116.13\deg\ & $-$32.01\deg\  & 0.58\arcsec\ & 576 $\times$ 384 &  15 Jun -- 16 Jul & 2.30\arcsec \nl
\enddata
\tablenotetext{a}{Median Seeing for observations reported here.}
\tablenotetext{b}{Scattered additional observations were also performed 
in September and October.}
\end{deluxetable}


\begin{deluxetable}{lrrrrcr}
\tablewidth{440.0pt}
\tablenum{2} \label{frames}
\tablecaption{Number of Reduced Frames}
\tablehead{
\colhead{Event\tablenotemark{a}} & \multicolumn{2}{r}{La~Silla 0.9m} 
 & \multicolumn{2}{r}{SAAO 1.0m} & \colhead{Perth 0.6m} & \colhead{Total}\\
 & I & V & I & V & I & 
}
\startdata
MACHO 95-BLG-10 & ~~~~ 109 & 37 & ~~~~ 48 & 14 & & 208 \nl
MACHO 95-BLG-12 & 122 & 32 & 78 & 38 & 26 & 296 \nl
MACHO 95-BLG-13 & 152 & 41 & 78 & 59 & 21 & 351 \nl
MACHO 95-BLG-17 &  93 & 24 & 10 &    &   & 127 \nl
MACHO 95-BLG-18 &  65 & 13 & 23 &  6 &   & 107 \nl
MACHO 95-BLG-19 &  72 & 21 & 56 & 18 &   & 167 \nl
MACHO 95-BLG-21 &     &    & 11 &    &   &  11 \nl
MACHO 95-BLG-24 &  13 &  6 &  9 &  5 &   &  33 \nl
MACHO 95-BLG-30 &  29 & 25 &    &    &   &  54 \nl
OGLE~ 95-BLG-04 &  78 & 28 & 23 &  7 &   & 136 \nl
\enddata
\tablenotetext{a}{The name of the event, as designated by the discovery 
team, is given in the first column.}
\end{deluxetable}


\begin{deluxetable}{lllllrrr}
\tablewidth{470.0pt}
\tablenum{3} \label{machotable}
\tablecaption{MACHO- and OGLE-Reported Event Parameters}
\tablehead{
\colhead{Event} & \colhead{RA (J2000)} & \colhead{Dec (J2000)}  
& \colhead{V\tablenotemark{a}} & \colhead{R\tablenotemark{a}} 
& \colhead{$A_{max}$} & \colhead{$t_o$\tablenotemark{b}} 
& \colhead{$t_E$\tablenotemark{c}} 
}
\startdata  
MACHO 95-BLG-10 & 17:58:16.0 & $-$29:32:11 & 18.9 & 18.0 &  2.8 & 151 & 47.5\nl
MACHO 95-BLG-12 & 18:06:04.8 & $-$29:52:38 & 18.6 & 17.7 & 3.16 & 155 & binary \nl
MACHO 95-BLG-13 & 18:08:47.0 & $-$27:40:47 & 16.6 & 15.6 &  3.8 & 179 & 73.5\nl
MACHO 95-BLG-17 & 18:03:01.1 & $-$28:21:09 & 18.8 & 18.0 &  1.9 & 163 & 18.5\nl 
MACHO 95-BLG-18 & 18:07:20.6 & $-$28:36:51 & 18.7 & 17.8 &  1.6 & 190 & 39.5\nl 
MACHO 95-BLG-19 & 18:11:32.5 & $-$27:45:27 & 18.6 & 17.9 &  4.2 & 187 & 31.5\nl 
MACHO 95-BLG-21 & 17:59:42.2 & $-$28:08:42 & 20.7 & 19.7 &  3.0 & 183 & 11.0\nl 
MACHO 95-BLG-24 & 18:02:54.4 & $-$29:26:30 & 17.8 & 16.9 &  4.5 & 186 &  7.0\nl
MACHO 95-BLG-30 & 18:07:04.3 & $-$27:22:06 & 16.1 & 14.7 & 25.0 & 226 & 33.9\nl
OGLE~ 95-BLG-04 & 18:02:07.6 & $-$30:01:13 & 20.00 & 18.45\tablenotemark{a} & 1.8 & 165 & 28.0\nl
\enddata
\tablecomments{Taken from OGLE alert notifications and the MACHO alert page at 
http://www.darkstar.astro.washington.edu.}
\tablenotetext{a}{MACHO broad band V and R magnitudes are quoted, except 
in the case of OGLE~ 95-BLG-04 for which the V and I magnitudes 
determined by the OGLE team are given.  Uncertainty in MACHO magnitudes 
are typically $\sim$10\%.}
\tablenotetext{b}{Approximate JD - 2449718.5, uncertainty typically 
on the order of 1-2 days.}
\tablenotetext{c}{Einstein radius crossing time in days.}
\end{deluxetable}


\begin{deluxetable}{lrrrrrr}
\tablewidth{480.0pt}
\tablenum{4} \label{planettable}
\tablecaption{PLANET-Derived Event Parameters}
\tablehead{
\colhead{Event} & \colhead{I\tablenotemark{a}} 
& \colhead{V-I\tablenotemark{b}} & \colhead{$A_{max}$} & 
\colhead{$t_o$\tablenotemark{c}} & \colhead{$t_E~\rm{(days)}$} 
}
\startdata
MACHO 95-BLG-10  & 17.5 (0.3) & 1.59 (0.11) & 3.3 (0.6) & 151\tablenotemark{d}  & 49 (24) \nl
MACHO 95-BLG-12  & 17.03 (0.14) & 1.63 (0.04) & ---  & --- & --- \nl
MACHO 95-BLG-13  & 14.75 (0.04)  & 1.85 (0.03) & 3.95 (0.12) & 180.3 (0.5) & 76.2 (5.0) \nl
MACHO 95-BLG-17  & 17.00 (0.04)  & 1.48 (0.04) & 1.61 (0.16) & 163\tablenotemark{d} & 8.9 (1.5) \nl
MACHO 95-BLG-18  & 17.16 (0.08) & 1.75 (0.07) & 2.5 (0.2) & 204 (6) & 61.5 (7.0) \nl
MACHO 95-BLG-19  & 17.62 (0.32) & 1.12 (0.04) & 4.93 (1.4) & 187.9 (0.21) & 35.4 (12.3) \nl
MACHO 95-BLG-21  & --- & ---  & 2.83 (0.28) & 184.5 (0.16) & 7.4 (2.40)\nl
MACHO 95-BLG-24  & 15.8 (0.08) & 1.64 (0.06) & --- & 186\tablenotemark{d} & --- \nl
MACHO 95-BLG-30  & 13.08 (0.08) & 2.97 (0.02)  & 25\tablenotemark{d} & 226\tablenotemark{d} & 35.7 (4.7) \nl
OGLE~ 95-BLG-04  & 17.9 (0.37) & 1.40 (0.13) &  1.36 (0.45) &  170.9 (4.8) & 11 (15) \nl
\enddata
\tablecomments{Colors and magnitudes have not been de-reddened; quoted uncertainties do not include an approximate 14\% uncertainty from 
transformation to the standard system.}
\tablenotetext{a}{Baseline Cousins I magnitudes are determined from fits.}
\tablenotetext{b}{$V-I$ color is determined from common fit to data unless 
otherwise noted.}
\tablenotetext{c}{HJD - 2449718.5}
\tablenotetext{d}{Held fixed at MACHO-determined value}.
\end{deluxetable}


\end{document}